\documentclass[12pt, draftclsnofoot, onecolumn]{IEEEtran}

\ifCLASSINFOpdf
\else
\fi

\usepackage{graphicx,amssymb,amsmath}
\usepackage{multicol}
\usepackage[noadjust]{cite}
\usepackage{setspace}
\usepackage{stfloats}
\usepackage{cite}
\usepackage{amsthm}
\usepackage{flushend}
\usepackage{cases,subeqnarray}
\usepackage{bm,multirow,bigstrut}
\usepackage{textcomp}
\usepackage{latexsym,bm}
\usepackage{xcolor}
\usepackage{mathtools}
\usepackage{dsfont}
\usepackage{extarrows}
\usepackage{algorithm}
\usepackage{algpseudocode}
\usepackage{amsmath}
\usepackage{booktabs}

\theoremstyle{plain}
\newtheorem{thm}{Theorem}
\newtheorem{lemm}{Lemma}
\theoremstyle{plain}
\newtheorem{rem}{Remark}
\newtheorem{coro}{Corollary}

\begin{document}

\title{Local Partial Zero-Forcing Combining for Cell-Free Massive MIMO Systems}

\author{Jiayi Zhang,~\IEEEmembership{Senior Member,~IEEE,}
        Jing Zhang,
        Emil Bj{\"o}rnson,~\IEEEmembership{Senior Member,~IEEE,}
        and Bo Ai,~\IEEEmembership{Senior Member,~IEEE}% <-this % stops a space
\thanks{This work was supported in part by National Key R\&D Program of China under Grant 2020YFB1807201, in part by National Natural Science Foundation of China under Grants 61971027, U1834210, and 61961130391, in part by Beijing Natural Science Foundation under Grant L202013, in part by Frontiers Science Center for Smart High-speed Railway System, in part by the Royal Society Newton Advanced Fellowship under Grant NA191006, in part by the Fundamental Research Funds for the Central Universities, China, under grant number 2020JBZD005, in part by the Project of China Shenhua under grant number (GJNY-20-01-1). E.~Bj\"ornson was supported by the Grant 2019-05068 from the Swedish Research Council. (\emph{Corresponding author: Jiayi Zhang.})}
\thanks{J. Zhang and J. Zhang are with the School of Electronic and Information Engineering, Beijing Jiaotong University, Beijing 100044, China, and also with the Frontiers Science Center for Smart High-speed Railway System, Beijing Jiaotong University, Beijing 100044, China (e-mail: jiayizhang@bjtu.edu.cn).}% <-this % stops a space
\thanks{E. Bj{\"o}rnson is with the Department of Electrical Engineering, Link{\"o}ping
University, Link{\"o}ping, Sweden, and the Department of Computer Science,
KTH Royal Institute of Technology, Kista, Sweden (e-mail: emilbjo@kth.se).}% <-this % stops a space
\thanks{B. Ai is with the State Key Laboratory of Rail Traffic Control and Safety, Beijing Jiaotong University, Beijing 100044, China, and also with the Frontiers Science Center for Smart High-speed Railway System, and also with Henan Joint International Research Laboratory of Intelligent Networking and Data Analysis, Zhengzhou University, Zhengzhou 450001, China, and also with Research Center of Networks and Communications, Peng Cheng Laboratory, Shenzhen, China (e-mail: boai@bjtu.edu.cn).}}

%% The paper headers

\maketitle
\vspace{-1cm}
\begin{abstract}
Cell-free massive multiple-input multiple-output (MIMO) provides more uniform spectral efficiency (SE) for users (UEs) than cellular technology. The main challenge to achieve the benefits of cell-free massive MIMO is to realize signal processing in a scalable way. In this paper, we consider scalable full-pilot zero-forcing (FZF), partial FZF (PFZF), protective weak PFZF (PWPFZF), and local regularized ZF (LRZF) combining by exploiting channel statistics. We derive closed-form expressions of the uplink SE for FZF, PFZF, and PWPFZF combining with large-scale fading decoding over independent Rayleigh fading channels, taking channel estimation errors and pilot contamination into account. Moreover, we investigate the impact of the number of pilot sequences, antennas per AP, and APs on the performance. Numerical results show that LRZF provides the highest SE. However, PWPFZF is preferable when the number of pilot sequences is large and the number of antennas per AP is small. The reason is that PWPFZF has lower computational complexity and the SE expression can be computed in closed-form. Furthermore, we investigate the performance of PWPFZF combining with fractional power control and the numerical results show that it improves the performance of weak UEs and realizes uniformly good service for all UEs in a scalable fashion.
\end{abstract}


% \fi
\IEEEpeerreviewmaketitle

\section{Introduction}
The cellular network deployments have been utilized to support the rapid data traffic growth, which has made inter-cell interference the major bottleneck \cite{zhang2020Prospective,Lopez-Perez2011Enhanced,ngo2017on,Andrews2016Are, mai2020downlink}.
Network multiple-input multiple-output (MIMO) technology can suppress such interference through joint coherent cooperation between access points (APs) \cite{venkatesan2007network}.
In particular, a higher spectral efficiency (SE) is obtained than when each user equipment (UE) is served by only one selected AP \cite{ngo2017on,nayebi2017precoding,bjornson2019making}.
However, to achieve the excellent theoretical gains of the early network MIMO methods, network-wide channel state information (CSI) must be gathered at the APs \cite{venkatesan2007network,gesbert2010multi-cell}.
This is impractical due to the immense fronthaul signaling and huge computational complexity, which poses performance limitations and system scalability issues \cite{bjornson2019scalable}.
Cell-free massive MIMO is a more practical embodiment of the network MIMO concept \cite{ngo2017cell, nayebi2017precoding, papazafeiropoulos2020towards,Zhang2019cell}, where CSI is not shared between the APs. Conceptually, it is a time-division duplex (TDD) distributed massive MIMO system with a large number of APs that coherently serve all the UEs on the same time-frequency resource \cite{ngo2017cell}.
Each AP is connected to a central processing unit (CPU) which is responsible to coordinate and process the signals of UEs \cite{ngo2017cell, nayebi2017precoding,bjornson2019making}.
Cell-free massive MIMO is different from conventional distributed antenna systems, where the antennas are distributed within each cell.
In contrast, there are no cells in cell-free massive MIMO and all service antennas coherently serve all UEs \cite{ngo2017cell}.
The basic concept of having APs connected to CPUs resembles the cloud radio access network (C-RAN) architecture.
However, C-RAN is based on dividing the APs in a cellular network into disjoint clusters, each connected to a separate CPU, and only permits cooperation within each cluster.
This is not a cell-free network since every cluster becomes a cell with distributed antennas.
In contrast, cell-free massive MIMO is a user-centric network where each UE is served by the most preferable set of APs \cite{Buzzi2018User,bjornson2019scalable}.
In this way, the cell concept vanishes and more uniform SEs can be delivered in the network.
The peak rates in cell-free deployment might reduce compared to cellular deployments, but there is a much higher chance of achieving a decent rate with a 95\% probability \cite{ngo2017cell,bjornson2019making}.
This is the main motivation behind cell-free massive MIMO.

Two outstanding aspects of cell-free massive MIMO are the large macro-diversity and favorable propagation which are facilitated by a large number of APs \cite{ngo2017cell}.
This has motivated the use of traditional maximum ratio (MR) processing \cite{ngo2017cell,nayebi2017precoding}, which is optimal when the inter-user interference is negligible.
However, in practical setups, the SE can be greatly improved by minimum mean-square error (MMSE) processing methods that actively suppress inter-user interference \cite{bjornson2019making,bjornson2019scalable}.
The highest SE is achieved in a centralized implementation where the CSI is sent to the CPU to enable joint interference suppression between the APs, but there are practical reasons for carrying out the receive combining and transmit precoding locally at every AP.

\textcolor{black}{Several prior works consider zero-forcing (ZF) processing in cell-free massive MIMO systems, e.g.,  \cite{[C1],[C2],[C4],interdonato2020local,[C3]}.
Centralized ZF schemes are studied in \cite{[C1],[C2],[C4]}, in which the instantaneous CSI needs to be sent from all APs to the CPU for designing the ZF precoding/combining vectors centrally.
%However, it is the centralized ZF scheme that is considered in \cite{[C1],[C2],[C4]}, which needs to be calculated at the CPU with instantaneous CSI being sent from all APs to the CPU.
However, this will result in unmanageable fronthauling traffic and an unscalable architecture when the number of UEs grows.
This motivates us to investigate the distributed schemes in this work where the processing is carried out at the APs.
Several distributed ZF precoding schemes, e.g., full-pilot ZF (FZF), partial FZF (PFZF), and protective PFZF (PPFZF) have been introduced in \cite{interdonato2020local}. All of them can suppress interference fully distributively or coordinately in a scalable fashion,
%The study in \cite{interdonato2020local} shows that these ZF precoding schemes can suppress the inter-user interference efficiently compared with MR precoding.
and shown to outperform the MR scheme.
Besides, the local regularized ZF (LRZF) is also investigated in \cite{interdonato2020local} and as a upper benchmark with the price of being non-scalable.
The performance of FZF and modified LRZF precoder are also studied in \cite{[C3]} considering NOMA-aided cell-free massive MIMO system.
However, the uplink was not considered in \cite{interdonato2020local} and \cite{[C3]}.}

Motivated by the above discussion, we investigate the uplink SE provided by MR, FZF, PFZF, and LRZF combining.
This requires a substantially different analysis compared with \cite{interdonato2020local}, since in the uplink, every AP computes a local estimate of the data signals and then the large-scale fading decoding (LSFD) scheme must be used to properly fuse these estimates at the CPU \cite{bjornson2019making}, \cite{nayebi2016performance}.
Besides, in order to improve the service quality of weak UEs, which is the main advantage of cell-free massive MIMO compared with cellular systems, we can alternatively apply the protective weak PFZF (PWPFZF) combining for weak UEs to significantly reduce the intra-group interference.
A key difference between uplink and downlink is which UEs can benefit from interference suppression.
In the uplink, it is only the UE that is assigned to the combining vector that benefits.
In the downlink, it is only other UEs than the one that is assigned to the precoding vector that can benefit.
Hence, different from \cite{interdonato2020local}, our PWPFZF aims at providing protection to weak UEs instead of strong UEs, although the protections are both realized by forcing the MR combining to take place in the orthogonal complement of the effective channels of the strong UEs.

Our contributions are listed as follows. First, using the use-and-then-forget (UatF) technique, we derive new closed-form uplink SE expressions with the FZF, PFZF, and PWPFZF combining schemes.
The expressions take LSFD, imperfect CSI, and pilot contamination into account.
The asymptotic closed-form SE expression with LRZF is also derived.
Then, we compare the uplink performance of MR, FZF, PFZF, PWPFZF, and LRZF with full power transmission in a cell-free massive MIMO system with different configurations.
Our results show that LRZF provides the highest SE while MR gives the lowest SE.
Although PWPFZF achieves a lower SE than LRZF, it is a good choice when the number of pilot sequences is large and the number of antennas per AP is small, since it has lower computational complexity and we can compute the SE in exact closed-form.
The performance of FZF, PFZF, and PWPFZF depends on the system parameters.
When the number of pilot sequences is small or the number of antennas per APs is large, FZF performs better.
Besides, in order to make all UEs have almost uniformly good service in a scalable fashion, we apply the fractional power control method proposed in \cite{nikbakht2019uplink} along with PWPFZF to further improve the 95\%-likely SE per SE.

The most closely related work is \cite{interdonato2020local}, which considers FZF, PFZF, and PWPFZF precoding and per-AP power control, which are two separable problems in the downlink.
In this paper, we consider the uplink where the power control is done on a per-UE basis, but there is instead the need for utilizing the LSFD scheme in the uplink to properly weigh the inputs from the different APs together at the CPU.
As combining schemes, we consider the uplink counterparts of FZF, PFZF, and PWPFZF and use the same naming convention, for simplicity, but we stress that the analysis becomes substantially different.
We derive new closed-form SE expressions and optimal LSFD weights.
We stress that new expressions cannot be obtained from \cite{interdonato2020local} using reciprocity arguments since \cite{interdonato2020local} considers fully distributed processing, while we consider LSFD processing.
Another related work is \cite{bjornson2019making}, which also considers the uplink but for two other combining schemes (MR and local minimum mean square error (local-MMSE), thus there is no overlap in terms of analytical contributions.
The proposed combining schemes outperform MR and provide comparable performance to local-MMSE, but have the added benefit of leading to closed-form expressions.
These expressions enable carrying out resource allocation tasks without having to first compute expectations by Monte Carlo methods. This will greatly simplify the practical implementation and algorithmic design.
Besides, using the derived closed-form expressions, we can understand how the system works, including what happens to the array gains when applying different types of interference suppression, how does the interference terms depend on the system parameters, etc.

\section{System Model}\label{system_model}
We analyze a cell-free massive MIMO system in which the single-antenna UEs are jointly served by all the APs. Let $L$, $N$, $K$ be the numbers of APs, antennas per AP, and UEs, respectively.
The APs are connected to a CPU via fronthaul links.
We consider the standard massive MIMO protocol from \cite{bjornson2017massive}, where each coherence block is divided into ${\tau _p}$ channel uses for uplink pilots and ${\tau _u}$ for uplink data such that ${\tau _c} = {\tau _p} + {\tau _u}$.
We use the block-fading model where ${{\bf{h}}_{kl}} \in {{\mathbb{C}}^{N \times 1}}$ is the channel response between the $k$th UE and the $l$th AP, $k = 1, \ldots ,K$, $l = 1, \ldots ,L$.
In each block, an independent realization from an independent Rayleigh fading distribution is drawn:
\begin{equation}
\textcolor{black}{
{{\bf{h}}_{kl}} \sim {{\cal N}_{\mathbb{C}}}\left( {\mathbf{0}},{{\beta _{kl}}{{\bf{I}}_N}} \right),}
\end{equation}
where ${{\beta _{kl}}}$ is the large-scale fading coefficient that describes geometric pathloss and shadowing. We assume the large-scale fading coefficients $\left\{ {{\beta _{kl}}} \right\}$ are available wherever needed in the network.

\subsection{Uplink Training and Channel Estimation}
In a pilot-based uplink training, all the UEs synchronously send their pilot sequences to the APs, once per coherence block.
We assume that there are ${\tau _p} \le K$ mutually orthogonal ${\tau _p}$-length pilot signals utilized and assigned to the UEs.
We let ${i_k} \in \left\{ {1, \ldots ,{\tau _p}} \right\}$ denote the index of the pilot used by UE $k$, therefore \textcolor{black}{${{\bm{\phi}} _{{i_k}}} \in {{\mathbb{C}}^{{\tau _p} \times 1}}$} is the pilot sequence sent by the $k$th UE.
We call ${{\cal P}_k} \subset \left\{ {1, \ldots ,K} \right\}$ the subset of UEs that use the same pilot as UE $k$, including itself, hence ${i_k} = {i_t} \Leftrightarrow t \in {{\cal P}_k}$.
Any two pilot sequences are orthogonal such that
\begin{equation}
{\pmb{\phi}} _{{i_t}}^H{{\pmb{\phi}} _{{i_k}}} = \left\{ {\begin{array}{*{20}{c}}
{0,}&{t \notin {{\cal P}_k},}\\
{{\tau _p},}&{t \in {{\cal P}_k}.}
\end{array}} \right.
\end{equation}
The received signal ${{\bf{Z}}_l} \in {{\mathbb{C}}^{N \times {\tau _p}}}$ at AP $l$ is given by
\begin{equation}
{{\bf{Z}}_l} = \sum\limits_{k = 1}^K {\sqrt {p_k^{\rm{p}}} } {{\bf{h}}_{kl}}{\bm{\phi}} _{{i_k}}^H + {{\bf{N}}_l},
\end{equation}
where ${{\bf{N}}_l} \in {{\mathbb{C}}^{N \times {\tau _p}}}$ is a Gaussian noise matrix with i.i.d. {${{\cal N}_{\mathbb{C}}}\left( {0,{\sigma^2}} \right)$} elements and ${p_k^{\rm{p}}}$ is the transmit power of UE $k$ for uplink training.
The MMSE estimate of ${{\bf{h}}_{kl}}$ given ${{\bf{Z}}_l}$ is \cite{Kay1996Fundamentals}
\begin{equation}
{{{\bf{\hat h}}}_{kl}} \buildrel \Delta \over = \frac{{{c_{kl}}}}{{\sqrt {{\tau _p}} }}{{\bf{Z}}_l}{{\bm{\phi}} _{{i_k}}},
\end{equation}
where
\begin{equation}
{c_{kl}} \buildrel \Delta \over = \frac{{{\sqrt {p_k^{\rm{p}}{\tau _p}} } {\beta _{kl}}}}{{{\tau _p}\sum\limits_{t \in {{\cal P}_k}} {p_t^{\rm{p}}} {\beta _{kl}} + {\sigma^2}}}.
\end{equation}
It can be verified that the estimate ${{{\bf{\hat h}}}_{kl}}$ and estimation error ${{{\bf{\tilde h}}}_{kl}} = {{\bf{h}}_{kl}} - {{{\bf{\hat h}}}_{kl}}$ are independent Gaussian vectors with distributions ${{{\bf{\hat h}}}_{kl}} \sim {{\cal N}_{\mathbb{C}}}\left( {{{0}},{\gamma _{kl}}{{\bf{I}}_N}} \right)$ and ${{{\bf{\tilde h}}}_{kl}} \sim {{\cal N}_{\mathbb{C}}}\left( {{{0}},\left( {{\beta _{kl}} - {\gamma _{kl}}} \right){{\bf{I}}_N}} \right)$, respectively,
where
\begin{equation}
{\gamma _{kl}} \buildrel \Delta \over = {\mathbb{E}}\left\{ {{{\left| {{{\left[ {{{{\bf{\hat h}}}_{kl}}} \right]}_n}} \right|}^2}}  \right\} = \frac{{p_k^{\rm{p}}{\tau _p}\beta _{kl}^2}}{{{\tau _p}\sum\limits_{t \in {{\cal P}_k}} {p_t^{\rm{p}}} {\beta _{tl}} + {\sigma ^2}}},\;n = 1 \ldots ,N.
\end{equation}

\begin{rem}
Note that the channel estimates are parallel when the pilot sequences assigned to two different UEs are the same.
In particular, when UEs $k$ and $t$, $t \ne k$, use the same pilot, the channel estimates ${{\mathbf{h}}_{kl}}$ and ${{\mathbf{h}}_{tl}}$ are linearly dependent, $l = 1, \cdots ,L$,
\begin{equation}
{{{\bf{\hat h}}}_{kl}} = \frac{{\sqrt {p_k^{\rm{p}}} {\beta _{kl}}}}{{\sqrt {p_t^{\rm{p}}} {\beta _{tl}}}}{{{\bf{\hat h}}}_{tl}}.
\end{equation}
Hence, the AP cannot separate the UEs sharing the same pilot and cannot suppress the corresponding pilot contamination interference.
\end{rem}

\subsection{Uplink Data Transmission}
During the data transmission, the received complex baseband signal ${{\bf{y}}_l} \in {{\mathbb{C}}^{N \times 1}}$ at AP $l$ is
\begin{equation}\label{uplnk_signal}
{{\bf{y}}_l} = \sum\limits_{k = 1}^K {{{\bf{h}}_{kl}}} {s_k} + {{\bf{n}}_l},
\end{equation}
where ${s_k} \sim {{\cal N}_{\mathbb{C}}}\left( {0,p_k^{{\rm{ul}}}} \right)$ is the information-bearing signal transmitted by UE $k$ with power $p_k^{{\rm{ul}}}$ and ${{\bf{n}}_l} \sim {{\cal N}_{\mathbb{C}}}\left( {0,{\sigma ^2}{{\bf{I}}_N}} \right)$ is the independent receiver noise.

Based on the signals in (\ref{uplnk_signal}), the APs and CPU decode the symbols with the LSFD technique.
The general idea of LSFD is that each AP computes local estimates of the desired data of all UEs in the first layer and transmits these to the CPU for final decoding in the second layer.
In detail, an estimate of the data symbol from UE $k$ at AP $l$ is obtained by local linear combining using the vector ${{\bf{v}}_{kl}} \in {{\mathbb{C}}^{N \times 1}}$ in the first layer as
\begin{align}\label{s_kl}
{\hat s_{kl}} = {\bf{v}}_{kl}^H{{\bf{y}}_l} &= {\bf{v}}_{kl}^H{{\bf{h}}_{kl}}{s_k} + {\bf{v}}_{kl}^H\sum\limits_{t \in {{\cal P}_k}/\left\{ k \right\}}^K {{{\bf{h}}_{tl}}} {s_t} + {\bf{v}}_{kl}^H\sum\limits_{t \notin {{\cal P}_k}}^K {{{\bf{h}}_{tl}}} {s_t} + {\bf{v}}_{kl}^H{{\bf{n}}_l}.
\end{align}
After the local data estimation, the second layer of centralized decoding is performed based on the local estimates $\left\{ {{{\hat s}_{kl}}:l = 1, \ldots ,L} \right\}$ using the LSFD coefficients $\left\{ {{a_{kl}}:l = 1, \ldots ,L} \right\}$ to obtain ${{\hat s}_k} = \sum\limits_{l = 1}^L {a_{kl}^*} {{\hat s}_{kl}}$.
From (\ref{s_kl}), we have that
\begin{align}\label{s_k}
{\hat s_k} &= \sum\limits_{l = 1}^L {a_{kl}^*} {\bf{v}}_{kl}^H{{\bf{h}}_{kl}}{s_k} + \sum\limits_{l = 1}^L {a_{kl}^*} {\bf{v}}_{kl}^H\sum\limits_{t \in {{\cal P}_k}/\left\{ k \right\}}^K {{{\bf{h}}_{tl}}} {s_t}+ \sum\limits_{l = 1}^L {a_{kl}^*} {\bf{v}}_{kl}^H\sum\limits_{t \notin {{\cal P}_k}}^K {{{\bf{h}}_{tl}}} {s_t} + \sum\limits_{l = 1}^L {a_{kl}^*} {\bf{v}}_{kl}^H{{\bf{n}}_l}.
\end{align}

As pointed out in Remark 1, when ${\tau _p} < K$, so that multiple UEs are assigned to each pilot, thus ${{\bf{\hat H}}_l} = \left[ {{{{\bf{\hat h}}}_{1l}}, \ldots ,{{{\bf{\hat h}}}_{Kl}}} \right] \in {{\mathbb{C}}^{N \times K}}$ is rank-deficient.
We can construct the full-rank matrix of the channel estimates, ${{{\bf{\bar H}}}_l}$, as
\begin{equation}
{{\bf{\bar H}}_l} = {{\bf{Z}}_l}{\bm{\Phi}}  \in {{\mathbb{C}}^{N \times {\tau _p}}}.
\end{equation}
Therefore, the channel estimate ${{{\bf{\hat h}}}_{kl}}$ can be expressed as
\begin{equation}
{{{\bf{\hat h}}}_{kl}} = {c_{kl}}{{{\bf{\bar H}}}_l}{{\bf{e}}_{{i_k}}},
\end{equation}
where ${\bm{\Phi}}  = \left[ {{{\pmb{\phi}} _1}, \ldots ,{{\pmb{\phi}} _{{\tau _p}}}} \right] \in {{\mathbb{C}}^{{\tau _p} \times {\tau _p}}}$ and ${{\bf{e}}_{{i_k}}}$ denotes the ${i_k}$th column of ${{\bf{I}}_{{\tau _p}}}$.

\section{Performance Analysis}
In this section, we derive and analyze achievable uplink {SE} expressions for different ZF-based combining schemes with LSFD in the considered cell-free massive MIMO system.

\subsection{Uplink Spectral Efficiency}
A standard capacity lower bound, i.e., an achievable SE, can be derived by utilizing the bounding technique in \cite[Sec. 2.3.2]{Marzetta2016Fundamentals}, \cite{Medard2000effect}, and \cite{bjornson2017massive}.
The expression in (\ref{s_k}) can be rewritten as
\begin{align}\label{E}
{{\hat s}_k} &= {s_k}\underbrace {\sum\limits_{l = 1}^L {a_{kl}^*} \sqrt {p_k^{{\rm{ul}}}} {\mathbb{E}}\left\{ {{\bf{v}}_{kl}^H{{\bf{h}}_{kl}}} \right\}}_{{\rm{D}}{{\rm{S}}_k}}
+ {s_k}\underbrace {\sum\limits_{l = 1}^L {a_{kl}^*} \sqrt {p_k^{{\rm{ul}}}} \left( {{\bf{v}}_{kl}^H{{\bf{h}}_{kl}} - {\mathbb{E}}\left\{ {{\bf{v}}_{kl}^H{{\bf{h}}_{kl}}} \right\}} \right)}_{{\rm{B}}{{\rm{U}}_k}}\notag\\
&+ \sum\limits_{t \in {{\cal P}_k}/\left\{ k \right\}}^K {{s_t}\underbrace {\sum\limits_{l = 1}^L {a_{kl}^*} \sqrt {p_t^{{\rm{ul}}}} {\bf{v}}_{kl}^H{{\bf{h}}_{tl}}}_{{\rm{P}}{{\rm{C}}_{kt}}}}
+ \sum\limits_{t \notin {{\cal P}_k}}^K {{s_t}\underbrace {\sum\limits_{l = 1}^L {a_{kl}^*} \sqrt {p_t^{{\rm{ul}}}} {\bf{v}}_{kl}^H{{\bf{h}}_{tl}}}_{{\rm{U}}{{\rm{I}}_{kt}}}}  + \underbrace {\sum\limits_{l = 1}^L {a_{kl}^*} {\bf{v}}_{kl}^H{{\bf{n}}_l}}_{{n_k}},
\end{align}
where ${\rm{D}}{{\rm{S}}_k}$, ${\rm{B}}{{\rm{U}}_k}$, ${{\rm{P}}{{\rm{C}}_{kt}}}$, ${{\rm{U}}{{\rm{I}}_{kt}}}$, and ${n_k}$ reflect the coherent beamforming gain, beamforming gain uncertainty, pilot contamination, inter-user interference, and noise, respectively.

By invoking the arguments from \cite{bjornson2019making}, the achievable uplink SE for UE $k$, can be written as stated in Lemma 1.
\begin{lemm}
Using LSFD, an achievable SE of UE $k$ is
\begin{equation}
{\rm{S}}{{\rm{E}}_k} = \left( {1 - \frac{{{\tau _p}}}{{{\tau _c}}}} \right){\log _2}\left( {1 + {\rm{SIN}}{{\rm{R}}_k}} \right)
\end{equation}
with the effective ${{\rm{SINR}}}$ is given by
\begin{align}\label{SINR}
{\rm{SIN}}{{\rm{R}}_k} & =\frac{{\mathbb{E}}{{{\left| {{\rm{D}}{{\rm{S}}_k}} \right|}^2}}}{{{\mathbb{E}}\left\{ {{{\left| {{\rm{B}}{{\rm{U}}_k}} \right|}^2}} \right\} + \sum\limits_{t \in {{\cal P}_k}/\left\{ k \right\}}^K {\mathbb{E}}{\left\{ {{{\left| {{\rm{P}}{{\rm{C}}_{kt}}} \right|}^2}} \right\}}  + \sum\limits_{t \notin {{\cal P}_k}}^K {\mathbb{E}}{\left\{ {{{\left| {{\rm{U}}{{\rm{I}}_{kt}}} \right|}^2}} \right\}}  +  {\mathbb{E}}\left\{ {{{\left| {{n_k}} \right|}^2}} \right\}}}\notag\\
&= \frac{{p_k^{{\rm{ul}}}{{\left| {{\bf{a}}_k^H{\mathbb{E}}\left\{ {{{\bf{g}}_{kk}}} \right\}} \right|}^2}}}{{\sum\limits_{i = 1}^K {p_i^{{\rm{ul}}}{\mathbb{E}}\left\{ {{{\left| {{\bf{a}}_k^H{{\bf{g}}_{ki}}} \right|}^2}} \right\}}  - p_k^{{\rm{ul}}}{{\left| {{\bf{a}}_k^H{\mathbb{E}}\left\{ {{{\bf{g}}_{kk}}} \right\}} \right|}^2} + {\sigma ^2}{\bf{a}}_k^H{{\bf{F}}_k}{{\bf{a}}_k}}},
\end{align}
where ${{\bf{g}}_{ki}} \!=\! {\left[ {{\bf{v}}_{k1}^H{{\bf{h}}_{i1}}, \ldots ,{\bf{v}}_{kL}^H{{\bf{h}}_{iL}}} \right]^T}$, ${{\bf{F}}_k} \!= \! {\rm{diag}}\left( {\left\{ {{{\left\| {{{\bf{v}}_{k1}}} \right\|}^2}} \right\}, \ldots ,\left\{ {{{\left\| {{{\bf{v}}_{kL}}} \right\|}^2}} \right\}} \right)$,
${{\bf{a}}_k} \!= \!{\left[ {{a_{k1}}, \ldots ,{a_{kL}}} \right]^T}$.
The effective SINR in (\ref{SINR}) can be further maximized by
\begin{equation}
{{\bf{a}}_k} = {\left( {\sum\limits_{t = 1}^K {p_t^{{\rm{ul}}}\left\{ {{{\bf{g}}_{kt}}{\bf{g}}_{kt}^H} \right\}}  + {\sigma ^2}{{\bf{F}}_k}} \right)^{ - 1}}\left\{ {{{\bf{g}}_{kk}}} \right\}
\end{equation}
which leads to the maximum value
\begin{align}\label{SINR_max}
&{\rm{SINR}}_k^{{\rm{max}}} = p_k^{{\rm{ul}}}\left\{ {{\bf{g}}_{kk}^H} \right\} {\left( {\sum\limits_{t = 1}^K {p_i^{{\rm{ul}}}} \left\{ {{{\bf{g}}_{kt}}{\bf{g}}_{kt}^H} \right\} + {\sigma ^2}{{\bf{F}}_k} - p_t^{{\rm{ul}}}\left\{ {{{\bf{g}}_{kt}}} \right\}} \right)^{ - 1}}\left\{ {{{\bf{g}}_{kk}}} \right\}.
\end{align}
\end{lemm}

\subsection{Maximum {Ratio} Combining}
The simplest solution is maximum ratio (MR) combining with
\begin{equation}\label{MR}
{{\bf{v}}_{kl}} = {{{\bf{\hat h}}}_{kl}} = {c_{kl}}{{{\bf{\bar H}}}_l}{{\bf{e}}_{{i_k}}},
\end{equation}
which has low computational complexity and maximizes the power of the desired signal in the numerator of (\ref{SINR}).
By plugging (\ref{MR}) into (\ref{SINR_max}), and calculating the corresponding expected values, the achievable uplink SE can be obtained in closed-form for MR combining in \cite{Zhang2020Cell-Free}.

\subsection{Full-pilot Zero-Forcing Combining}
Despite the low complexity, MR combining is known to be a vastly suboptimal scheme {\cite{bjornson2019making}}, since it neglects the existence of interference in the denominator of (\ref{SINR}).
The centralized ZF combining has been employed to suppress the inter-user interference in cell-free massive MIMO and analyzed in several works under the assumption of no pilot contamination \cite{nayebi2017precoding} \cite{ngo2017on}.
However, implementing centralized ZF combining requires instantaneous channel state information (CSI) to be sent from the APs to the CPU to construct combining vectors.

Unlike the centralized ZF combining, FZF combining has the ability to suppress interference in a fully distributed, coordinated, and scalable fashion, which means that the APs do not send the instantaneous CSI to the CPU as the combining vectors are constructed at the APs.
Note that the local nature of this combination is extremely important to preserve the system scalability.
Besides, the computation of FZF combining has much lower complexity than the centralized ZF, which is resulting from the matrix inversion in FZF combining has a much lower dimension.
Finally, FZF combining can be analyzed in the presence of pilot contamination and give insightful closed-form SE expressions.

The local combining vector that AP $l$ selects for UE $k$, ${\bf{v}}_{{i_k}l}^{{\rm{FZF}}} \in {{\mathbb{C}}^{N \times 1}}$, is given by
\begin{equation}\label{V_FZF}
{\bf{v}}_{{i_k}l}^{{\rm{FZF}}} = {c_{{i_k}l}}{\theta _{{i_k}l}}{{{\bf{\bar H}}}_l}{\left( {{\bf{\bar H}}_l^H{{{\bf{\bar H}}}_l}} \right)^{ - 1}}{{\bf{e}}_{{i_k}}}.
\end{equation}
\begin{rem}
Each AP has ${\tau _p}$ combining vectors, one per pilot.
Therefore, the capability to cancel interference is highly dependent on the number of AP antennas, which must meet the requirement $N  \ge  {\tau _p}$.
The same vector is used for all the UEs sharing the same pilot.
\end{rem}
As mentioned in Remark 2, the APs cannot distinguish the UEs that share the same pilot, therefore, employing FZF combining can suppress the interference towards the UEs that use different pilots:
\begin{align}\label{FZF}
&{\mathbf{v}}_{{i_k}l}^H{{{\mathbf{\hat h}}}_{tl}} = {c_{{i_t}l}}{c_{{i_k}l}}{\theta _{{i_k}l}}{\left( {{{\mathbf{e}}_{{i_k}}}} \right)^H}{\left( {{\mathbf{\bar H}}_l^H{{{\mathbf{\bar H}}}_l}} \right)^{ - 1}}{\mathbf{\bar H}}_l^H{{{\mathbf{\bar H}}}_l}{{\mathbf{e}}_{{i_t}}}= {c_{{i_t}l}}{c_{{i_k}l}}{\theta _{{i_k}l}}{\mathbf{e}}_{{i_k}}^H{{\mathbf{e}}_{{i_t}}} = \left\{ {\begin{array}{*{20}{c}}
  {0,t \notin {\mathcal{P}_k},} \\
  {{\gamma _{{i_k}l}},t \in {\mathcal{P}_k},}
\end{array}} \right.
\end{align}
where ${\theta _{{i_k}l}} = {\mathbb{E}}\left\{ {{{\left| {{{\left[ {{{{\bf{\bar H}}}_l}{{\bf{e}}_{{i_k}}}} \right]}_n}} \right|}^2}} \right\} = \frac{{{\gamma _{{i_k}l}}}}{{c_{{i_k}l}^2}}$.
By substituting (\ref{FZF}) into (\ref{SINR_max}), and computing the expected values, the ergodic SE is obtained in closed-form.
\begin{coro}
A lower bound on the uplink ergodic capacity in Lemma 1, for i.i.d. Rayleigh fading channels and FZF combining, is given by
\begin{equation}\label{SE_FZF}
{\rm{SE}}_k^{{\rm{FZF}}} = \left( {1 - \frac{{{\tau _p}}}{{{\tau _c}}}} \right){\log _2}\left( {1 + {\rm{SINR}}_k^{{\rm{FZF}}}} \right),
\end{equation}
where ${{\rm{SINR}}_k^{{\rm{FZF}}}}$ is given as
\begin{equation}\label{SINR_FZF}
{\rm{SINR}}_k^{{\rm{FZF}}} = \frac{{\left( {N - {\tau _p}} \right)p_k^{{\rm{ul}}}{{\left| {\sum\limits_{l = 1}^L {a_{kl}^*} {\gamma _{kl}}} \right|}^2}}}{{\sum\limits_{t = 1}^K {p_t^{{\rm{ul}}}\sum\limits_{l = 1}^L {{{\left| {a_{kl}^*} \right|}^2}} {\gamma _{kl}}\left( {{\beta _{tl}} - {\gamma _{tl}}} \right)}  + \left( {N - {\tau _p}} \right)\sum\limits_{t \in {\mathcal{P}_k}/\left\{ k \right\}}^K {p_t^{{\rm{ul}}}{{\left( {\sum\limits_{l = 1}^L {a_{kl}^*{\gamma _{tl}}} } \right)}^2}}  + {\sigma ^2}\sum\limits_{l = 1}^L {{{\left| {a_{kl}^*} \right|}^2}{\gamma _{kl}}} }}.
\end{equation}

The optimal LSFD vector can also be obtained in closed-form as ${{\bf{a}}_k} = {\bf{C}}_k^{ - 1}{{\bf{b}}_k}$, where ${{\mathbf{b}}_k} = {\left[ {{\gamma _{k1}}, \ldots ,{\gamma _{kL}}} \right]^H}$ and
\begin{align}
{{\mathbf{C}}_k} &= \sum\limits_{t \in {\mathcal{P}_k}/\left\{ k \right\}}^K {p_t^{{\rm{ul}}}{{\mathbf{b}}_t}{\mathbf{b}}_t^H}  +  {\rm{diag}}\left( {\sum\limits_{i = 1}^K {p_t^{{\rm{ul}}}} {\gamma _{k1}}\left( {{\beta _{t1}} - {\gamma _{t1}}} \right) +  {\sigma ^2}{\gamma _{k1}}}, \right.\notag\\
&\left. { \cdots ,\sum\limits_{i = 1}^K {p_t^{{\rm{ul}}}} {\gamma _{kL}}\left( {{\beta _{tL}} - {\gamma _{tL}}} \right) + {\sigma ^2}{\gamma _{kL}}} \right).
\end{align}

Then, with FZF combining, (\ref{SE_FZF}) is given in the closed-form as
\begin{equation}
{\rm{SE}}_k^{{\rm{FZF}}} = \left( {1 - \frac{{{\tau _p}}}{{{\tau _c}}}} \right){\log _2}\left( {1 + p_k^{{\rm{ul}}}{\bf{b}}_k^H{\bf{C}}_k^{ - 1}{{\bf{b}}_k}} \right).
\end{equation}
\end{coro}
\begin{IEEEproof}
Please see Appendix \ref{proofFZF}.
\end{IEEEproof}

Compared with MR combining, when the weights $a_{kl}, \forall k,l$ are the same, part of the interference, i.e., the first term of the denominator of (\ref{SINR_FZF}), can be significantly reduced from ${{\beta _{tl}}}$ to ${{\beta _{tl}} - {\gamma _{tl}}}$ for UE $k$ at AP $l$ with no additional fronthauling overhead.
The remain coherent interference, (i.e., the second term of the denominator of (\ref{SINR_FZF})), also becomes smaller.
The cost is a loss in array gain of ${{\tau _p}}$, since the array gain with FZF combining is ${\left( {N - {\tau _p}} \right)}$.

\subsection{Partial Full-pilot Zero Forcing Combining}
FZF combining spends ${{\tau _p}}$ degrees of freedom to cancel the interference.
However, the inter-user interference that affects UE $k$ is mainly generated by a small subset of the other UEs.
Inspired by this, we apply the PFZF combining that only suppresses the interference generated by strong UEs that have strong channel gains.
Conversely, the interference generated by weak UEs which have weak channel gains is tolerated.
Therefore, AP $l$ employs the PFZF combining for strong UEs and MR combining for weak UEs.
The principle of using some antenna for suppressing the interference and the others for boosting the desired signal has already been applied in \cite{Jindal2011Multi}, \cite{Veetil2015Performance} for MIMO communication, in \cite{Fang2017Coverage} for millimeter-wave cellular network and in \cite{Buzzi2018User} for cell-free massive MIMO.

We adopt the similar grouping principle as in \cite{interdonato2020local}.
At AP $l$, all the UEs are divided into two groups: ${{\cal S}_l} \subset \left\{ {1, \ldots ,K} \right\}$ gathers strong UEs while ${{\cal W}_l} \subset \left\{ {1, \ldots ,K} \right\}$ gathers weak UEs.
The UE $k$ belongs to ${{\cal S}_l}$ if ${\beta _{kl}}$ is above a predetermined threshold, else UE $k$ belongs to ${{\cal W}_l}$.

\begin{rem}
APs cannot distinguish UEs that share the same pilot, therefore, these UEs are assigned to the same group.
\end{rem}

Since only strong UEs use the PFZF combining, we define ${\tau _{{{\cal S}_l}}}$ as the number of different pilots used by the UEs ${ \in {{\cal S}_l}}$ and ${{\cal R}_{{{\cal S}_l}}} = \left( {{r_{l,1}}, \ldots ,{r_{l,{\tau _{{{\cal S}_l}}}}}} \right)$ as the set of the corresponding pilot indices.
Therefore, the pilot-book matrix for UEs ${ \in {{\cal S}_l}}$ is given by {${{\bm{\Phi}} _{{{\cal S}_l}}}{\rm{ = }}{\bm{\Phi }} {{\bf{E}}_{{{\cal S}_l}}}$,} where ${{\bf{E}}_{{{\cal S}_l}}} = \left( {{{\bf{e}}_{{r_{l,1}}}}, \ldots ,{{\bf{e}}_{{r_{l,{\tau _{{{\cal S}_l}}}}}}}} \right) \in {{\mathbb{C}}^{{\tau _p} \times {\tau _{{{\cal S}_l}}}}}$ and ${{{\bf{e}}_{{r_{l,t}}}}}$ is the ${{r_{l,t}}}$th column of ${{\bf{I}}_{{\tau _p}}}$.
With respect to ${{\bm{\Phi}} _{{{\cal S}_l}}}$, we define ${j_{kl}} \in \left\{ {1, \ldots ,{\tau _{{{\cal S}_l}}}} \right\}$ the index.
Let {${{\bm{\varepsilon}} _{{j_{kl}}}} \in {{\mathbb{C}}^{{\tau _{{{\cal S}_l}}} \times 1}}$} as the ${j_{kl}}$th column of ${{\bf{I}}_{{\tau _{{{\cal S}_l}}}}}$, and it leads to ${{\bf{E}}_{{{\cal S}_l}}}{\varepsilon _{{j_{kl}}}}{\rm{ = }}{{\bf{e}}_{{i_k}}}$.
Then, the PFZF combining for UE $k$ $\in {{\cal S}_l}$ at AP $l$ is given as
\begin{equation}\label{V_PFZF}
{{\bf{v}}_{{i_k}l}^{{\rm{PFZF}}}} = {c_{{i_k}l}}{\theta _{{i_k}l}}{{{\bf{\bar H}}}_l}{{\bf{E}}_{{{\cal S}_l}}}{\left( {{\bf{E}}_{{{\cal S}_l}}^H{\bf{\bar H}}_l^H{{{\bf{\bar H}}}_l}{{\bf{E}}_{{{\cal S}_l}}}} \right)^{ - 1}}{\varepsilon _{{j_{kl}}}}.
\end{equation}
\begin{rem}
At AP $l$, if all the UEs are assigned into ${{\cal S}_l}$, then ${{\cal S}_l} = \left\{ {1, \ldots ,K} \right\}$, ${\tau _{{{\cal S}_l}}} = {\tau _p}$, ${{\bf{E}}_{{{\cal S}_l}}} = {{\bf{I}}_{{\tau _p}}}$ and ${\varepsilon _{{j_{kl}}}} = {{\bf{e}}_{{i_k}}}$.
As a result, PFZF becomes identical to FZF.
\end{rem}

With PFZF combining for UEs $\in {{\cal S}_l}$ and MR combining for UEs $\in {{\cal W}_l}$, (\ref{s_k}) can be rewritten as
\begin{align}
&{{\hat s}_k} = \sum\limits_{l = 1}^L {a_{kl}^*} {{\hat s}_{kl}} = \sum\limits_{l \in {\mathcal{Z}_k}} {a_{kl}^*} \left( {{{\left( {{\mathbf{v}}_{kl}^{{\rm{FZF}}}} \right)}^H}{{\mathbf{h}}_{kl}}{s_k}+{{\left( {{\mathbf{v}}_{kl}^{{\rm{FZF}}}} \right)}^H}\sum\limits_{t \in {\mathcal{W}_l}} {{{\mathbf{h}}_{tl}}{s_t}} + {\left( {{\mathbf{v}}_{kl}^{{\rm{FZF}}}} \right)^H}\sum\limits_{t \in {\mathcal{P}_{k,{\mathcal{S}_l}}}/\left\{ k \right\}}  {{{\mathbf{h}}_{tl}}{s_t}}} \right.\notag\\
&  + {\left( {{\mathbf{v}}_{kl}^{{\rm{FZF}}}} \right)^H}\sum\limits_{t \notin {\mathcal{P}_{k,{\mathcal{S}_l}}}} {{{\mathbf{h}}_{tl}}{s_t}} \left. {+ {{\left( {{\mathbf{v}}_{kl}^{{\rm{FZF}}}} \right)}^H}{{\mathbf{n}}_l}} \right)
+\sum\limits_{l \in {\mathcal{M}_k}} {a_{kl}^*} \left( {{{\left( {{\mathbf{v}}_{kl}^{{\rm{MR}}}} \right)}^H}{{\mathbf{h}}_{kl}}{s_k}+{{\left( {{\mathbf{v}}_{kl}^{{\rm{MR}}}} \right)}^H}\sum\limits_{t \in {\mathcal{S}_l}} {{{\mathbf{h}}_{tl}}{s_t}} } \right. \notag\\
& \left. { + {\left( {{\mathbf{v}}_{kl}^{{\rm{MR}}}} \right)^H}{{\mathbf{n}}_l} + {{\left( {{\mathbf{v}}_{kl}^{{\rm{MR}}}} \right)}^H}\sum\limits_{t \in {\mathcal{P}_{k,{\mathcal{W}_l}}}/\left\{ k \right\}} {{{\mathbf{h}}_{tl}}{s_t}}  + {{\left( {{\mathbf{v}}_{kl}^{{\rm{MR}}}} \right)}^H}\sum\limits_{t \notin {\mathcal{P}_{k,{\mathcal{W}_l}}}} {{{\mathbf{h}}_{tl}}{s_t}} } \right).
\end{align}
where ${{\cal P}_k} \subset \left\{ {1, \ldots ,\left| {{{\cal S}_l}} \right|} \right\}$ refers to the set of UEs in ${{\cal S}_l}$ that use the same pilot as UE $k$, ${{\cal Z}_k} = \left\{ {l = 1, \ldots ,L:k \in {{\cal S}_l}} \right\}$ refers to the set of APs that assign UE $k$ into strong UEs and ${{\cal M}_k} = \left\{ {l = 1, \ldots ,L:k \in {{\cal W}_l}} \right\}$ refers to the set of APs that assign UE $k$ into weak UEs.
Since AP $l$ only use the PFZF combining for UEs $\in {{\cal S}_l}$ and MR combining is still employed for UEs $\in {{\cal W}_l}$, the intra-group interference between UEs $\in {{\cal S}_l}$ is actively suppressed, while the inter-group interference between UEs $\in {{\cal W}_l}$ and UEs $\in {{\cal S}_l}$ is tolerated.
Hence, for any pair of UEs $k$, $t \in {{\cal S}_l}$
\begin{align}\label{V_PFZF_S}
{{\left( {{\bf{v}}_{{i_k}l}^{{\rm{FZF}}}} \right)}^H}{{{\bf{\hat h}}}_{tl}}&= {c_{{i_t}l}}{c_{{i_k}l}}{\theta _{{i_k}l}}\varepsilon _{{j_{kl}}}^H{\left( {{\bf{E}}_{{{\cal S}_l}}^H{\bf{\bar H}}_l^H{{{\bf{\bar H}}}_l}{{\bf{E}}_{{{\cal S}_l}}}} \right)^{ - 1}}{\bf{E}}_{{{\cal S}_l}}^H{\bf{\bar H}}_l^H{{{\bf{\bar H}}}_l}{{\bf{E}}_{{{\cal S}_l}}}{\varepsilon _{{j_{tl}}}}= \left\{ {\begin{array}{*{20}{c}}
{0,\;\;\;t \notin {{\cal P}_k},}\\
{{\gamma _{{i_k}l}},\;\;t \in {{\cal P}_k}.}
\end{array}} \right.
\end{align}
For any pair of UEs $k$, $t \in {{\cal W}_l}$
\begin{equation}\label{V_PFZF_W}
{\mathbb{E}}\left\{ {{{\left( {{\bf{v}}_{kl}^{{\rm{MR}}}} \right)}^H}{{\bf{h}}_{tl}}} \right\} = {\mathbb{E}}\left\{ {{\bf{\hat h}}_{kl}^H{{\bf{h}}_{tl}}} \right\} = \left\{ {\begin{array}{*{20}{c}}
{0,\;\;\;\;\;t \notin {{\cal P}_k},}\\
{N{\gamma _{kl}},t \in {{\cal P}_k},}
\end{array}} \right.
\end{equation}

\begin{thm}
At any AP $l$, if PFZF combining is used for UEs $\in {{\cal S}_l}$ and MR combining is used for UEs $\in {{\cal W}_l}$, an achievable SE of UE $k$ is
\begin{equation}\label{SE_PFZF}
{\rm{SE}}_k^{{\rm{PFZF}}} = \left( {1 - \frac{{{\tau _p}}}{{{\tau _c}}}} \right){\log _2}\left( {1 + {\rm{SINR}}_k^{{\rm{PFZF}}}} \right),
\end{equation}
with the effective SINR given by
\begin{align}\label{SINR_PFZF}
&{\rm{SINR}}_k^{{\rm{PFZF}}} =\frac{{p_k^{{\rm{ul}}}{{\left| {{\bf{x}}_k^H{\mathbb{E}}\left\{ {{{\bf{y}}_{kk}}} \right\}} \right|}^2}}}{{\sum\limits_{t = 1}^K {p_t^{{\rm{ul}}}{\mathbb{E}}\left\{ {{{\left| {{\bf{x}}_k^H{{\bf{y}}_{kt}}} \right|}^2}} \right\}}  - p_k^{{\rm{ul}}}{{\left| {{\bf{x}}_k^H{\mathbb{E}}\left\{ {{{\bf{y}}_{kk}}} \right\}} \right|}^2} + {\sigma ^2}{\bf{a}}_k^H{{\bf{W}}_k}{{\bf{a}}_k}}},
\end{align}
where
\begin{align}
&{{\bf{x}}_k} = {\left[ {{a_{k{l_{{z_1}}}}}, \ldots ,{a_{kl{z_{\left| {{{\cal Z}_k}} \right|}}}},{a_{k{l_{{m_1}}}}}, \ldots ,{a_{k{l_{{m_{\left| {{{\cal M}_k}} \right|}}}}}}} \right]^T},\notag\\
&{{\mathbf{y}}_{kt}} = \left[ {{{\left( {{\mathbf{v}}_{k{l_{{z_1}}}}^{{\rm{FZF}}}} \right)}^H}{{\mathbf{h}}_{t{l_{{z_1}}}}}, \ldots ,{{\left( {{\mathbf{v}}_{kl{z_{\left| {{\mathcal{Z}_k}} \right|}}}^{{\rm{FZF}}}} \right)}^H}{{\mathbf{h}}_{tl{z_{\left| {{\mathcal{Z}_k}} \right|}}}},} \right.{\left. {{{\left( {{\mathbf{v}}_{k{l_{{m_1}}}}^{{\rm{MR}}}} \right)}^H}{{\mathbf{h}}_{t{l_{{m_1}}}}}, \ldots ,{{\left( {{\mathbf{v}}_{k{l_{{m_{\left| {{\mathcal{M}_k}} \right|}}}}}^{{\rm{MR}}}} \right)}^H}{{\mathbf{h}}_{t{l_{{m_{\left| {{\mathcal{M}_k}} \right|}}}}}}} \right]^T},\notag\\
&{{\mathbf{z}}_k} = \left[ {{{\left( {{\mathbf{v}}_{k{l_{{z_1}}}}^{{\rm{FZF}}}} \right)}^H}{{\mathbf{n}}_{{\mathbf{y}}_{{l_{{z_1}}}}^{{\rm{ul}}}}}, \ldots ,{{\left( {{\mathbf{v}}_{k{l_{{z_1}}}}^{{\rm{FZF}}}} \right)}^H}{{\mathbf{n}}_{{\mathbf{y}}_{_{{z_{\left| {{\mathcal{Z}_k}} \right|}}}}^{{\rm{ul}}}}},} \right.{\left. {{{\left( {{\mathbf{v}}_{k{l_{{m_1}}}}^{{\rm{MR}}}} \right)}^H}{{\mathbf{n}}_{{\mathbf{y}}_{{l_{_{{m_1}}}}}^{{\rm{ul}}}}}, \ldots ,{{\left( {{\mathbf{v}}_{k{l_{{m_{\left| {{\mathcal{M}_k}} \right|}}}}}^{{\rm{MR}}}} \right)}^H}{{\mathbf{n}}_{{\mathbf{y}}_{{l_{_{{m_{\left| {{\mathcal{M}_k}} \right|}}}}}}^{{\rm{ul}}}}}} \right]^T},\notag\\
&{{\mathbf{W}}_k} = {\rm{diag}}\left( {{{\left\| {{\mathbf{v}}_{k{l_{{z_1}}}}^{{\rm{FZF}}}} \right\|}^2}, \ldots ,{{\left\| {{\mathbf{v}}_{kl{z_{\left| {{\mathcal{Z}_k}} \right|}}}^{{\rm{FZF}}}} \right\|}^2},} \right.\left. {{{\left\| {{\mathbf{v}}_{k{l_{{m_1}}}}^{{\rm{MR}}}} \right\|}^2}, \ldots ,{{\left\| {{\mathbf{v}}_{k{l_{{m_{\left| {{\mathcal{M}_k}} \right|}}}}}^{{\rm{MR}}}} \right\|}^2}} \right).
\end{align}
The optimal LSFD vector ${{\bf{x}}_k}$ can be computed as
\begin{equation}
{{\bf{a}}_k} =  {\left( {\sum\limits_{t = 1}^K {p_t^{{\rm{ul}}}{\mathbb{E}}\left\{ {{{\bf{y}}_{kt}}{\bf{y}}_{kt}^H} \right\}}  + {\sigma ^2}{{\bf{w}}_k}} \right)^{ - 1}}{\mathbb{E}}\left\{ {{{\bf{y}}_{kk}}} \right\}.
\end{equation}
\end{thm}

\begin{coro}
By using PFZF combining for UEs $\in {{\cal S}_l}$ at AP $l$ and MR combining for UEs $\in {{\cal W}_l}$ at AP $l$, the SINR {of UE $k$} in (\ref{SINR_PFZF}) is given in the closed-form as
\begin{equation}
{\rm{SINR}}_k^{{\rm{PFZF}}} =  \frac{{p_k^{{\rm{ul}}}{{\left| {\sum\limits_{l \in {{\cal Z}_k}} {a_{kl}^*{\gamma _{kl}}}  + \sum\limits_{l \in {{\cal M}_k}} {Na_{kl}^*{\gamma _{kl}}} } \right|}^2}}}{{{\bf{Z}}_k^{{\rm{PFZF}}} + {\bf{M}}_k^{{\rm{PFZF}}} +\sum\nolimits_{t \in {{\cal P}_k}/\left\{ k \right\}}^K {{\bf{A}}_{kt}^{{\rm{PFZF}}}}  +  {\bf{Q}}_k^{{\rm{PFZF}}}}},
\end{equation}
where
\begin{align}
&{\bf{Z}}_k^{{\rm{PFZF}}} = \sum\limits_{t = 1}^K {p_t^{{\rm{ul}}}\sum\limits_{l \in {{\cal Z}_k}} {{{\left| {a_{kl}^*} \right|}^2}{\frac{{{\gamma _{kl}}\left( {{\beta _{tl}} - {\gamma _{tl}}} \right)}}{{\left( {N - {\tau _{{\mathcal{S}_l}}}} \right)}}}}},\;\;\;{\bf{M}}_k^{{\rm{PFZF}}} = {\sum\limits_{t = 1}^K {p_t^{{\rm{ul}}}\sum\limits_{l \in {\mathcal{M}_k}} {{{\left| {a_{kl}^*} \right|}^2}} N{\gamma _{kl}}{\beta _{tl}}} },\notag\\
&{\bf{A}}_{kt}^{{\rm{PFZF}}} = p_t^{{\rm{ul}}}{{{\left( {\sum\limits_{l \in {\mathcal{Z}_k}} {a_{kl}^*{\gamma _{tl}}}  +  N\sum\limits_{l \in {\mathcal{M}_k}} {a_{kl}^*{\gamma _{tl}}} } \right)}^2}},\notag\\
&{\bf{Q}}_k^{{\rm{PFZF}}} = {\sigma ^2}\sum\limits_{l \in {{\cal Z}_k}}{\frac{{{{\left| {a_{kl}^*} \right|}^2}{\gamma _{kl}}}}{{\left( {N - {\tau _{{\mathcal{S}_l}}}} \right)}}}  + {\sigma ^2}\sum\limits_{l \in {{\cal M}_k}} {{{\left| {a_{kl}^*} \right|}^2}} N{\gamma _{kl}}.
\end{align}

The optimal LSFD vector can also be obtained in closed-form as
${{\bf{a}}_k} = {\bf{C}}_k^{ - 1}{{\bf{b}}_k}$,
where
\begin{equation}
{{\bf{C}}_k} = \sum\limits_{t \in {{\cal P}_k}/\left\{ k \right\}}^K {p_t^{{\rm{ul}}}{{\bf{b}}_k}{\bf{b}}_k^H}  + {\rm{diag}}\left( {{\bf{W}}_{k1}^{{\rm{PFZF}}}, \cdots ,{\bf{W}}_{kL}^{{\rm{PFZF}}}} \right),
\end{equation}
where
\begin{equation}
{m_{kl}} = \left\{ {\begin{array}{*{20}{c}}
  {1,l \in {\mathcal{Z}_k}}, \\
  {0,l \notin {\mathcal{Z}_k}},
\end{array}} \right.\;\;\;
{n_{kl}} = \left\{ {\begin{array}{*{20}{c}}
  {1,l \in {\mathcal{M}_k}}, \\
  {0,l \notin {\mathcal{M}_k}},
\end{array}} \right.,
\end{equation}
\begin{align}
{{\bf{W}}_{kl}} &= \sum\limits_{t = 1}^K {p_t^{{\rm{ul}}}{\delta _{kl}}\frac{{{\gamma _{kl}}\left( {{\beta _{tl}} - {\gamma _{tl}}} \right)}}{{\left( {N - {\tau _{{{\cal S}_l}}}} \right)}}}  + \sum\limits_{t = 1}^K {p_t^{{\rm{ul}}}{n_{kl}}N{\gamma _{kl}}{\beta _{tl}}} {\sigma ^2}{m_{kl}}\frac{{{\gamma _{kl}}}}{{\left( {N - {\tau _{{{\cal S}_l}}}} \right)}} + {\sigma ^2}{n_{kl}}N{\gamma _{kl}},
\end{align}
and
\begin{align}
{{\mathbf{b}}_t} = {\left[ {{\gamma _{t{l_{{z_1}}}}}, \ldots ,{\gamma _{tl{z_{\left| {{\mathcal{Z}_k}} \right|}}}},N{\gamma _{t{l_{{m_1}}}}},N{\gamma _{t{l_{{m_{\left| {{\mathcal{M}_k}} \right|}}}}}}} \right]^H}.
\end{align}

Then, the SE in (\ref{SE_PFZF}) is given in the closed-form as
\begin{equation}
{\rm{SE}}_k^{{\rm{PFZF}}} = \left( {1 - \frac{{{\tau _p}}}{{{\tau _c}}}} \right){\log _2}\left( {1 + p_k^{{\rm{ul}}}{\bf{b}}_k^H{\bf{C}}_k^{ - 1}{{\bf{b}}_k}} \right).
\end{equation}
\end{coro}
\begin{IEEEproof}
Please see  Appendix \ref{proofPFZF}.
\end{IEEEproof}

By applying PFZF combining for strong UEs and MR combining for weak UEs, the inter-user interference can be suppressed if both UE $k$ and any UE $t$ are in the strong UE set, however, the pilot contamination still exists no matter which group UEs are in and the inter-user interference generated by weak UEs also exists.
Compared with MR and FZF combining, employing PFZF combining and MR combining for different groups of UEs lead to a balance between suppressing interference and obtaining large array gain.
PFZF combining only cancels the interference with the cost of ${\tau _{{{\cal S}_l}}}$ and hence take the advantage of a larger array gain than FZF combining.

\subsection{Protective Weak Partial Full-pilot Zero-Forcing}
As mentioned in the previous section, UEs in ${{\cal W}_l}$ experience both inter-group interference which includes pilot contamination, inter-user interference, and intra-group interference generated by UEs in ${{\cal S}_l}$.
In order to improve the service quality of weak UEs, which is the main advantage of cell-free massive MIMO compared with cellular systems, we can alternatively apply the PWPFZF combining for weak UEs to significantly reduce the intra-group interference.
The main idea of PWPFZF is to force the MR combining vector to take place in the orthogonal complement of ${{\bf{\bar H}}_l}{{\bf{E}}_{{{\cal S}_l}}}$, which are the effective channels of UEs in ${{\cal S}_l}$.
A key difference between uplink and downlink is which UEs can benefit from interference suppression.
In the uplink, it is only the UE that is assigned to the combining vector that benefits.
In the downlink, it is only other UEs than the one that is assigned to the precoding vector that can benefit.
Hence, different from \cite{interdonato2020local}, our PWPFZF aims at providing protection to weak UEs instead of strong UEs, although the protections are both realized by forcing the MR combining to take place in the orthogonal complement of the effective channels of the strong UEs, ${{\mathbf{\bar H}}_l}{{\mathbf{\bar E}}_{{\mathcal{S}_l}}}$.

With PWPFZF, the MR combining used at AP $l$ for UEs in ${{\cal W}_l}$ is now given by
\begin{equation}
{{\bf{v}}_{kl}^{{\rm{PMR}}} = {c_{kl}}{\theta _{{k}l}}{{\bf{B}}_l}{{{\bf{\bar H}}}_l}{{\bf{e}}_{{i_k}}}},
\end{equation}
where
\begin{equation}
{{\bf{B}}_l} = {{\bf{I}}_N} - {{{\bf{\bar H}}}_l}{{\bf{E}}_{{{\cal S}_l}}}{\left( {{\bf{E}}_{{{\cal S}_l}}^H{\bf{\bar H}}_l^H{{{\bf{\bar H}}}_l}{{\bf{E}}_{{{\cal S}_l}}}} \right)^{ - 1}}{\bf{E}}_{{{\cal S}_l}}^H{\bf{\bar H}}_l^H
\end{equation}
represents the projection matrix onto the orthogonal complement of ${{\bf{\bar H}}_l}{{\bf{E}}_{{{\cal S}_l}}}$.
Therefore, we have ${\left( {{\bf{v}}_{kl}^{{\rm{PMR}}}} \right)^H}{{\bf{h}}_{tl}} = 0$ if $t \in {{\cal S}_l}$.

\begin{thm}
By substituting ${{\bf{v}}_{kl}^{{\rm{PMR}}}}$ for ${{\bf{v}}_{kl}^{{\rm{MR}}}}$, {an} achievable SE for the PWPFZF scheme is given by
\begin{equation}\label{SE_PWPFZF}
{\rm{SE}}_k^{{\rm{PWPFZF}}} = \left( {1 - \frac{{{\tau _p}}}{{{\tau _c}}}} \right){\log _2}\left( {1 + {\rm{SINR}}_k^{{\rm{PFZF}}}} \right),
\end{equation}
with the effective SINR given by
\begin{align}\label{SINR_PWPFZF}
&{\rm{SINR}}_k^{{\rm{PWPFZF}}} = \frac{{p_k^{{\rm{ul}}}{{\left| {{\bf{x}}_k^H{\mathbb{E}}\left\{ {{{\bf{y}}_{kk}}} \right\}} \right|}^2}}}{{\sum\limits_{t = 1}^K  {p_t^{{\rm{ul}}}{\mathbb{E}}\left\{ {{{\left| {{\bf{x}}_k^H{{\bf{y}}_{kt}}} \right|}^2}} \right\}} - p_k^{{\rm{ul}}}{{\left| {{\bf{x}}_k^H{\mathbb{E}}\left\{ {{{\bf{y}}_{kk}}} \right\}} \right|}^2} + {\sigma ^2}{\bf{a}}_k^H{{\bf{W}}_k}{{\bf{a}}_k}}},
\end{align}
where
\begin{align}
&{{\bf{x}}_k} = {\left[ {{a_{k{l_{{z_1}}}}}, \ldots ,{a_{kl{z_{\left| {{{\cal Z}_k}} \right|}}}},{a_{k{l_{{m_1}}}}}, \ldots ,{a_{k{l_{{m_{\left| {{{\cal M}_k}} \right|}}}}}}} \right]^T},\notag\\
&{{\mathbf{y}}_{kt}} = \left[ {{{\left( {{\mathbf{v}}_{k{l_{{z_1}}}}^{{\rm{FZF}}}} \right)}^H}{{\mathbf{h}}_{t{l_{{z_1}}}}}, \ldots ,{{\left( {{\mathbf{v}}_{kl{z_{\left| {{\mathcal{Z}_k}} \right|}}}^{{\rm{FZF}}}} \right)}^H}{{\mathbf{h}}_{tl{z_{\left| {{\mathcal{Z}_k}} \right|}}}},} \right.
{\left. {{{\left( {{\mathbf{v}}_{k{l_{{m_1}}}}^{{\rm{PMR}}}} \right)}^H}{{\mathbf{h}}_{t{l_{{m_1}}}}}, \ldots ,{{\left( {{\mathbf{v}}_{k{l_{{m_{\left| {{\mathcal{M}_k}} \right|}}}}}^{{\rm{PMR}}}} \right)}^H}{{\mathbf{h}}_{t{l_{{m_{\left| {{\mathcal{M}_k}} \right|}}}}}}} \right]^T},\notag\\
&{{\mathbf{z}}_k} = \left[ {{{\left( {{\mathbf{v}}_{k{l_{{z_1}}}}^{{\rm{FZF}}}} \right)}^H}{{\mathbf{n}}_{{\mathbf{y}}_{{l_{{z_1}}}}^{{\rm{ul}}}}}, \ldots ,{{\left( {{\mathbf{v}}_{k{l_{{z_1}}}}^{{\rm{FZF}}}} \right)}^H}{{\mathbf{n}}_{{\mathbf{y}}_{_{{z_{\left| {{\mathcal{Z}_k}} \right|}}}}^{{\rm{ul}}}}},} \right.{\left. {{{\left( {{\mathbf{v}}_{k{l_{{m_1}}}}^{{\rm{PMR}}}} \right)}^H}{{\mathbf{n}}_{{\mathbf{y}}_{{l_{_{{m_1}}}}}^{{\rm{ul}}}}}, \ldots ,{{\left( {{\mathbf{v}}_{k{l_{{m_{\left| {{\mathcal{M}_k}} \right|}}}}}^{{\rm{PMR}}}} \right)}^H}{{\mathbf{n}}_{{\mathbf{y}}_{{l_{_{{m_{\left| {{\mathcal{M}_k}} \right|}}}}}}^{{\rm{ul}}}}}} \right]^T},\notag\\
&{{\mathbf{W}}_k} = {\rm{diag}}\left( {{{\left\| {{\mathbf{v}}_{k{l_{{z_1}}}}^{{\rm{FZF}}}} \right\|}^2}, \ldots ,{{\left\| {{\mathbf{v}}_{kl{z_{\left| {{\mathcal{Z}_k}} \right|}}}^{{\rm{FZF}}}} \right\|}^2},} \right.\left. {{{\left\| {{\mathbf{v}}_{k{l_{{m_1}}}}^{{\rm{PMR}}}} \right\|}^2}, \ldots ,{{\left\| {{\mathbf{v}}_{k{l_{{m_{\left| {{\mathcal{M}_k}} \right|}}}}}^{{\rm{PMR}}}} \right\|}^2}} \right).
\end{align}
The optimal LSFD vector ${{\bf{x}}_k}$ can be computed as
\begin{equation}
{{\bf{a}}_k} =  {\left( {\sum\limits_{t = 1}^K {p_t^{{\rm{ul}}}{\mathbb{E}}\left\{ {{{\bf{y}}_{kt}}{\bf{y}}_{kt}^H} \right\}}  + {\sigma ^2}{{\bf{w}}_k}} \right)^{ - 1}}{\mathbb{E}}\left\{ {{{\bf{y}}_{kk}}} \right\}.
\end{equation}
\end{thm}
\begin{coro}
By using PWPFZF combining for UEs $\in {{\cal S}_l}$ at AP $l$ and MR combining for UEs $\in {{\cal W}_l}$ at AP $l$, the SINR in (\ref{SINR_PWPFZF}) is given in the closed-form as
\begin{align}
&{\rm{SINR}}_k^{{\rm{PWPFZF}}} =\frac{p_k^{{\rm{ul}}}{\left| {\sum\limits_{l \in {{\cal Z}_k}} {a_{kl}^*{\gamma _{kl}}}  + \sum\limits_{l \in {{\cal M}_k}} {a_{kl}^*{\gamma _{kl}}\left( {N - {\tau _{{{\cal S}_l}}}} \right)} } \right|^2}}{{{\bf{Z}}_k^{{\rm{PWPFZF}}} +  {\bf{M}}_k^{{\rm{PWPFZF}}} + \sum\limits_{t \in {{\cal P}_k}/\left\{ k \right\}}^K {{\bf{A}}_{kt}^{{\rm{PWPFZF}}}}  +  {\bf{Q}}_k^{{\rm{PWPFZF}}}}},
\end{align}
where
\begin{align}
&{\bf{Z}}_k^{{\rm{PWPFZF}}} = \sum\limits_{t = 1}^K {p_t^{{\rm{ul}}}\sum\limits_{l \in {{\cal Z}_k}} {{{\left| {a_{kl}^*} \right|}^2}{\frac{{{\gamma _{kl}}\left( {{\beta _{tl}} - {\gamma _{tl}}} \right)}}{{\left( {N - {\tau _{{\mathcal{S}_l}}}} \right)}}}}}, \notag\\
&{\bf{M}}_k^{{\rm{PWPFZF}}} = \sum\limits_{t = 1}^K {p_t^{{\rm{ul}}}\sum\limits_{l \in {\mathcal{M}_k}} {{{\left| {a_{kl}^*} \right|}^2}} \left( {N - {\tau _{{\mathcal{S}_l}}}} \right){\gamma _{kl}}{\beta _{tl}}},\notag\\
&{\bf{A}}_{kt}^{{\rm{PWPFZF}}} = p_t^{{\rm{ul}}}{\left( {\sum\limits_{l \in {{\cal Z}_k}} {a_{kl}^*{{\gamma _{kl}}}}  + \sum\limits_{l \in {{\cal M}_k}} {a_{kl}^*{\gamma _{kl}}\left( {N - {\tau _{{{\cal S}_l}}}} \right)} } \right)^2},\notag\\
&{\bf{Q}}_k^{{\rm{PWPFZF}}} = {\sigma ^2}\sum\limits_{l \in {\mathcal{Z}_k}} {\frac{{{{\left| {a_{kl}^*} \right|}^2}{\gamma _{kl}}}}{{\left( {N - {\tau _{{\mathcal{S}_l}}}} \right)}}}  + {\sigma ^2}\sum\limits_{l \in {\mathcal{M}_k}}  {{{\left| {a_{kl}^*} \right|}^2}} \left( {N - {\tau _{{\mathcal{S}_l}}}} \right){\gamma _{kl}}.
\end{align}

The optimal LSFD vector can also be obtained in closed-form as
${{\bf{a}}_k} = {\bf{C}}_k^{ - 1}{{\bf{b}}_k}$,
where
\begin{equation}
{{\mathbf{C}}_k} = \sum\limits_{t \in {\mathcal{P}_k}/\left\{ k \right\}}^K  {p_t^{{\rm{ul}}}{{\mathbf{b}}_t}{\mathbf{b}}_t^H}  + {\rm{diag}}\left( {{\bf{W}}_{k1}^{{\rm{PWPFZF}}}, \cdots  ,{\bf{W}}_{kL}^{{\rm{PWPFZF}}}} \right),
\end{equation}
\begin{align}
{\bf{W}}_{kl}^{{\rm{PWPFZF}}} &= \sum\limits_{t = 1}^K {p_t^{{\rm{ul}}}{\delta _{kl}}\frac{{{\gamma _{kl}}\left( {{\beta _{tl}} - {\gamma _{tl}}} \right)}}{{\left( {N - {\tau _{{{\cal S}_l}}}} \right)}}}+ {\sigma ^2}{m_{kl}}\frac{{{\gamma _{kl}}}}{{\left( {N - {\tau _{{{\cal S}_l}}}} \right)}}\notag\\
&+ \sum\limits_{t = 1}^K {p_t^{{\rm{ul}}}{n_{kl}}\left( {N - {\tau _{{{\cal S}_l}}}} \right){\gamma _{kl}}{\beta _{tl}}} + {\sigma ^2}{n_{kl}}\left( {N - {\tau _{{{\cal S}_l}}}} \right){\gamma _{kl}},
\end{align}
and
\begin{align}
{{\bf{b}}_t} = {\left[ {{\gamma _{t{l_{{z_1}}}}}, \ldots ,{\gamma _{tl{z_{\left| {{\mathcal{Z}_k}}  \right|}}}},\left( {N - {\tau _{{\mathcal{S}_l}}}} \right){\gamma _{t{l_{{m_1}}}}},\left( {N - {\tau _{{\mathcal{S}_l}}}} \right){\gamma _{t{l_{{m_{\left| {{\mathcal{M}_k}} \right|}}}}}}} \right]^H}.
\end{align}

Then, the SE in (\ref{SE_PWPFZF}) is given in the closed-form as
\begin{equation}
{\rm{SE}}_k^{{\rm{PWPFZF}}} = \left( {1 - \frac{{{\tau _p}}}{{{\tau _c}}}} \right){\log _2}\left( {1 + p_k^{{\rm{ul}}}{\bf{b}}_k^H{\bf{C}}_k^{ - 1}{{\bf{b}}_k}} \right).
\end{equation}
\end{coro}
\begin{IEEEproof}
Please see  Appendix \ref{proofPWPFZF}.
\end{IEEEproof}

\subsection{Local Regularized Zero-Forcing}
As mentioned before, using PFZF combining for strong UEs and MR combining for weak UEs leads to a trade-off between interference suppression and boosting of the desired signal.
Similarly, the local regularized ZF (LRZF) combining scheme provides weighting between interference suppression and maximizing the intended signal \cite{bjornson2017massive} \cite{Peel2005vector}.

The LRZF combining for UE $k$ at AP $l$ is given by
\begin{equation}\label{V_LRZF}
{\bf{v}}_{kl}^{{\rm{LRZF}}} = {{{\bf{\hat H}}}_l}\left( {{\bf{\hat H}}_l^H{{{\bf{\hat H}}}_l} + {\sigma ^2}{\varphi _l}{{\bf{P}}^{ - 1}}} \right){{{\bf{\hat e}}}_k},
\end{equation}
where ${\varphi _l} = \sum\limits_{t = 1}^K {p_{tl}^{{\rm{ul}}}\left( {{\beta _{tl}} - {\gamma _{tl}}} \right)}$, ${\bf{P}} = {\rm{diag}}\left( {{p_1}, \ldots ,{p_K}} \right) \in {{\mathbb{C}}^{K \times K}}$, and ${{{\bf{\hat e}}}_k}$ is the $k$th column of ${{\bf{I}}_K}$.
However, it is intractable to derive a closed-form expression of the achievable SE due to the regularization term \cite{interdonato2020local}.

\begin{rem}
The LRZF combining vector in (\ref{V_LRZF}) involves the inversion of a $K$-dimensional matrix, but actually, the dimension can be reduced by some matrix algebra.
According to \cite{bjornson2017massive}, the matrix form of LRZF that gathers all $K$ combining vector can be written as
\begin{align}
{\bf{V}}_l^{{\rm{LRZF}}} &= \left[ {{\bf{v}}_{1l}^{{\rm{LRZF}}}, \ldots ,{\bf{v}}_{Kl}^{{\rm{LRZF}}}} \right]
= {\left( {{{{\bf{\hat H}}}_l}{\bf{P\hat H}}_l^H + {\sigma ^2}{\varphi _l}{{\bf{I}}_N}} \right)^{ - 1}}{{{\bf{\hat H}}}_l}{\bf{P}}\notag\\
&= {{{\bf{\bar H}}}_l}{\left( {{{\bf{F}}_l}{\bf{\bar H}}_l^H{{{\bf{\bar H}}}_l} + {\sigma ^2}{\varphi _l}{{\bf{I}}_{{\tau _p}}}} \right)^{ - 1}}\left[ {{p_1}{c_{1l}}{{\bf{e}}_{{t_1}}}, \ldots ,{p_K}{c_{Kl}}{{\bf{e}}_{{t_K}}}} \right],
\end{align}
where ${{\bf{F}}_l} = \sum\limits_{t = 1}^K {{p_t}c_{tl}^2{{\bf{e}}_{{i_t}}}{\bf{e}}_{{i_t}}^H} $.
Therefore, only a ${{{\tau _p}}}$-dimensional matrix needs to be inverted. Hence, the computational complexity with LRZF per AP in terms of number of complex multiplication is the same as FZF.
\end{rem}

Using LSFD, an achievable SE of UE $k$ with LRZF combining can be obtained from Theorem 1 by substituting (\ref{V_LRZF}) into (\ref{SINR_max}).
Besides, deriving the closed-form expression of the achievable SE with LRZF combining is too difficult because of the regularization term. Therefore, we only evaluate the performance of LRZF combining by using Lemma 1 and the corresponding expectations are computed by Monte-Carlo simulations.

\begin{figure}[t!]
\centering
\includegraphics[scale=0.75]{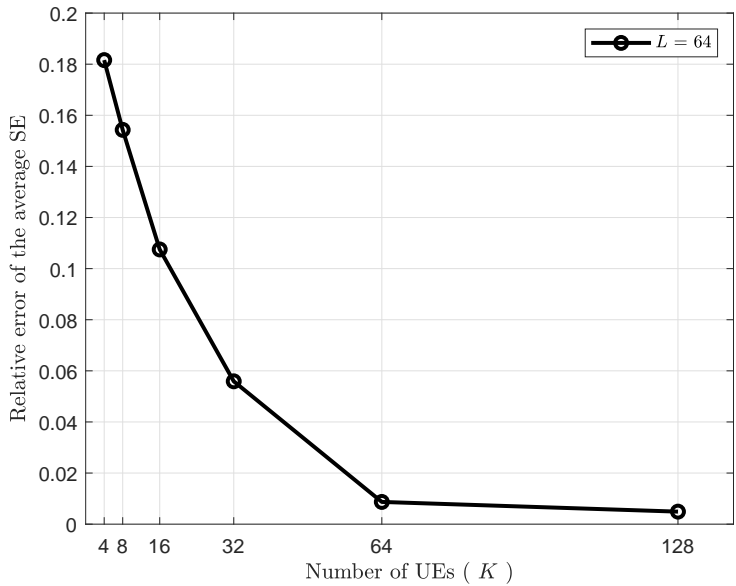}
\caption{The relative error of the average SE achieved by the asymptotic closed-form expression compared to the ergodic average SE versus the number of UEs, with $N = K$, $L = 64$, $\tau_p = K/2$, and $p_k = 100$ mW for each UE.}
\label{fig_mLRZF}
\end{figure}

\begin{rem}{
The modified LRZF (mLRZF) combining also tries to balance interference suppression and boosting of the desired signal.
Specifically, it is a special version of LRZF.
Since the closed-form analysis of mLRZF is difficult, we analyze its achievable SE in the asymptotic regime \cite{[C3],[R1],[R2]} when the number of antennas at each AP $N$ and the number of UEs $K$ grows large at the same pace.
Besides, the asymptotic analysis requires the following two assumptions: ${\lim _{N,K \to \infty }}{\sup _l}{\sup _{{\tau _k}}}{\theta _{{\tau _k}l}} < \infty $ and ${\lim _{N,K \to \infty }}{\inf _l}{\inf _{{\tau _k}}}{\theta _{{\tau _k}l}} > {\sigma ^2}$, where ${{\theta _{{\tau _k}l}}} =  {\sum\limits_{t \in {{\cal P}_k}} {p_t^{\rm{p}}{\tau _p}{\beta _{tl}}} + {\sigma ^2}} $ and ${\tau _k} \in \left\{ {1, \cdots ,{\tau _p}} \right\}$ represents the pilot index.
When $N$ and $K$ increase, the pilot assignment for existing UEs is fixed and the new UEs are assigned using some methods which keep the assumptions above.
The combining vector of mLRZF at the $l$th AP for UE $k$ can be expressed as
\begin{equation}
{{\bf{v}}_{{i_k}l}} = {c_{kl}}{\left( {{{{\bf{\bar H}}}_l}{\bf{\bar H}}_l^H + N\alpha {{\bf{I}}_N}} \right)^{ - 1}}{{{\bf{\bar h}}}_{{\tau _k}l}},
\end{equation}
where ${{{\bf{\bar H}}}_l}$ is given in (11), $\alpha  > 0$ is the regularization parameter, and ${\tau _k} \in \left\{ {1, \cdots ,{\tau _p}} \right\}$ represents the pilot index.
In the asymptotic regime, it holds that ${{\rm{SINR}}} _k - {{\rm{SINR}}}_k^{{\rm{mLRZF}}} \to 0$ where the asymptotic SINR expression is given as
\begin{equation}\label{SINR_mLRZF}
{\rm{SINR}}_k^{{\rm{mLRZF}}} = \frac{{p_k^{{\rm{ul}}}{{\left( {\sum\limits_{l = 1}^L {a_{kl}^*c_{kl}^2} \frac{{e_{{\tau _k}l}^ \circ }}{{1 + e_{{\tau _k}l}^ \circ }}} \right)}^2}}}{{\sum\limits_{t \in {{\cal P}_k}/\left\{ k \right\}}^K {p_t^{{\rm{ul}}}} {{\left( {\sum\limits_{l = 1}^L {a_{kl}^*{c_{kl}}{c_{tl}}\frac{{e_{{\tau _k}l}^ \circ }}{{1 + e_{{\tau _k}l}^ \circ }}} } \right)}^2} + \sum\limits_{l = 1}^L {{\Upsilon _{{\tau _k}}}\frac{{a_{kl}^2c_{kl}^2}}{{{{\left( {1 + {e_{{\tau _k}l}^ \circ }} \right)}^2}}}}  + {{\sigma ^2}\sum\limits_{l = 1}^L {a_{kl}^2c_{kl}^2\zeta _{{\tau _k}l}^ \circ } } }},
\end{equation}
where ${c_{kl}} = \frac{{\sqrt {p_k^{\rm{p}}{\tau _p}} {\beta _{kl}}}}{{\sum\nolimits_{t \in {{\cal P}_k}}^K {p_t^{\rm{p}}{\tau _p}{\beta _{tl}}}  + {\sigma ^2}}}$, ${{\theta _{{\tau _k}l}}} =  {\sum\limits_{t \in {{\cal P}_k}} {p_t^{\rm{p}}{\tau _p}{\beta _{tl}}}  + {\sigma ^2}}$, and $e_{{\tau _k}l}^ \circ  = {e_{{\tau _k}l}}$ in which
\begin{equation}\label{T}
{e_{{\tau _k}l}} = {{\theta _{{\tau _k}l}}}{{{T}}_l},\;\;\;\;\;\;\;\;\;\;
{{{T}}_l} = {\left( {\frac{1}{N}\sum\limits_{{\tau _i} = 1}^{{\tau _p}} {\frac{{{\theta _{{\tau _i}l}}}}{{1 + {e_{{\tau _i}l}}}} + \alpha } } \right)^{ - 1}},
\end{equation}
\begin{equation}
{e_{{\tau _i}{\tau _k}l}^{'}} = {{\theta _{{\tau _i}l}}}{{\theta _{{\tau _k}l}}}{T_l^{'}},
\end{equation}
\begin{equation}
{T_l^{'}} = {{{T}}_l}\left( {\frac{1}{N}\sum\limits_{{\tau _j} = 1}^{{\tau _p}} {\frac{{{{\theta _{{\tau _j}l}}}{e_{{\tau _j}l}^{'}}}}{{{{\left( {1 + {e_{{\tau _j}l}}} \right)}^2}}}}  + 1} \right){{{T}}_l},
\end{equation}
\begin{equation}
{e_{{\tau _k}l}^{'}} = {{\theta _{{\tau _k}l}}}{T_l^{'}},
\end{equation}
\begin{equation}
{\Upsilon _{{\tau _k}}} = \frac{1}{N}\sum\limits_{{\tau _i} \ne {\tau _k}}^K {p_i^{{\rm{ul}}}} \frac{{\left(  {c_{il}^2{e_{{\tau _i}{\tau _k}l}^{'}}{\theta _{{\tau _i}l}} + {{\left( {{\beta _{il}} - {\gamma _{il}}} \right)}}{e_{{\tau _k}l}^{'}}{{\left( {1 + {e_{{\tau _i}l}}} \right)}^2}} \right)}}{{{\theta _{{\tau _i}l}}{{\left( {1 + {e_{{\tau _i}l}}} \right)}^2}}},
\end{equation}
\begin{equation}
\zeta _{{\tau _k}l}  = \frac{1}{N}\frac{{{e_{{\tau _k}l}^{'}}}}{{{{\left( {1 + {e_{{\tau _k}l}}} \right)}^2}}},
\end{equation}
in which ${{{\bf{e}}}_l^{'}} = {\left[ {{{e}_{{\tau_1}l}^{'}}, \cdots ,{{e}_{{\tau_p}l}^{'}}} \right]^T}$ and ${{{\bf{e}}}_{{\tau_i}{\tau_k}l}^{'}} = {\left[ {{{e}_{{\tau_1}{\tau_k}l}^{'}}, \cdots ,{{e}_{{\tau_p}{\tau_k}l^{'}}}} \right]^T}$ are given by
\begin{align}
{{{\bf{e}}}_l^{'}} = {\left( {{{\bf{I}}_{{\tau _p}}} - {{\bf{J}}_l}} \right)^{ - 1}}{{\bf{w}}_l},\;\;\;
{{{\bf{e}}}_{{\tau _k}l}^{'}} = {\left( {{{\bf{I}}_{{\tau _p}}} - {{\bf{J}}_l}} \right)^{ - 1}}{{\bf{w}}_{{\tau _k}l}},
\end{align}
and ${{\bf{J}}_l}$, ${{\bf{w}}_l}$, and ${{\bf{w}}_{{\tau _k}l}}$ are derived as follows,
\begin{align}
&{\left[ {{{\bf{J}}_l}} \right]_{ij}} = \frac{{ {{{{\theta}} _{il}}{{{T}}_l}{{{\theta}} _{jl}}{{{T}}_l}} }}{{N{{\left( {1 + {e_{jl}}} \right)}^2}}},\notag\\
&{{\bf{w}}_l} = {\left[ { {{{{\theta}} _{{\tau _1}l}}{{T}}_l^2} , \cdots , {{{{\theta}} _{{\tau _p}l}}{{T}}_l^2} } \right]^T},\notag\\
&{{\bf{w}}_{{\tau _k}l}} = {\left[ { {{{{\theta}} _{{\tau _1}l}}{{{T}}_l}{{{\theta}} _{{\tau _k}l}}{{{T}}_l}} , \cdots , {{{{\theta}} _{{\tau _p}l}}{{{T}}_l}{{{\theta}} _{{\tau _k}l}}{{{T}}_l}} } \right]^T}.
\end{align}
Note that ${e_{{\tau _k}l}^ \circ }$ $\forall k$ is calculated using the fixed-point algorithm \cite{[R1]}.
Specifically, let $\left\{ {e_{{\tau _k}l}^{\left( i \right)}} \right\}\left( {i \ge 0} \right)$ $\forall k$ be the sequence defined by $e_{{\tau _k}l}^{0}  = \frac{1}{\alpha }$ $\forall k$ and $e_{{\tau _k}l}^{\left( i \right)} = {\theta _{{\tau _k}l}}{\left( {\frac{1}{N}\sum\limits_{{\tau _i} = 1}^{{\tau _p}} {\frac{{{\theta _{{\tau _k}l}}}}{{1 + e_{{\tau _i}l}^{\left( {i - 1} \right)}}}}  + \alpha } \right)^{ - 1}}$ for $i > 0$, then, ${\lim _{i \to \infty }}e_{{\tau _k}l}^{\left( i \right)} = {e_{{\tau _k}l}}$ $\forall k$.
Therefore, the initial value of ${e_{{\tau _k}l}}$ $\forall k$ to calculate (\ref{T}) are set to $e_{{\tau _k}l}^{0}  = \frac{1}{\alpha },{\tau _k} = 1, \cdots ,{\tau _p}$.
The final value of ${e_{{\tau _k}l}}$, ${e_{{\tau _k}l}^ \circ }$, is calculated after several iterations.
The simulation results presented in Fig. \ref{fig_mLRZF} depicts the error of the average SE
\begin{equation}
{\rm{SE}}_k^{{\rm{mLRZF}}} = \left( {1 - \frac{{{\tau _p}}}{{{\tau _c}}}} \right){\log _2}\left( {1 + {\rm{SINR}}_k^{{\rm{mLRZF}}}} \right),
\end{equation}
where ${\rm{SINR}}_k^{{\rm{mLRZF}}}$ is given in (\ref{SINR_mLRZF}), compared to the ergodic SE $\overline {{\rm{SE}}} _k^{{\rm{mLRZF}}}$ computed by Monte Carlo simulations.
The relative error of the average SE can be calculated as
\begin{equation}
E_k^{{\rm{mLRZF}}} = \left( {\overline {{\rm{SE}}} _k^{{\rm{mLRZF}}} - {\rm{SE}}_k^{{\rm{mLRZF}}}} \right)/\overline {{\rm{SE}}} _k^{{\rm{mLRZF}}}.
\end{equation}
Besides, we consider $\alpha = 0.8$.
Note that the value of $\alpha$ can be further optimized \cite{[C3]} and it will be  investigated in future work.
From Fig. \ref{fig_mLRZF}, we can observe that the approximated SE per UE becomes more accurate with increasing $N$ and $K$.
In particular, when $N = K = 128$, the relative error SE is 0.7\%.}
\end{rem}

\begin{rem}
The computational complexity with FZF, PFZF, PWPFZF, and LRZF combining schemes per AP in terms of number of complex multiplications can be derived follows form \cite{interdonato2020local}.
Among all four combining schemes, the computational complexity of LRZF and FZF are the highest.
Then, thanks to the fact that ${\tau _{{{\cal S}_l}}} \le {\tau _p}$, the complexity of PFZF and PWPFZF is lower than FZF and LRZF.
Besides, compared with PFZF, PWPFZF needs $2\left( {{\tau _p} - {\tau _{{{\cal S}_l}}}} \right){\tau _{{{\cal S}_l}}}N$ more complex multiplications for computing the ${{\tau _p} - {\tau _{{{\cal S}_l}}}}$ MR combining vectors in (\ref{V_PFZF_W}).
\end{rem}

\section{Numerical Results}
In this section, we compare the SEs provided by MR, FZF, PFZF, PWPFZF, and LRZF combining in cell-free massive MIMO.
The impact of the distribution of antennas and the number of pilot sequences are investigated based on the closed-form SE expression with LSFD for different combining schemes.
We adopt a similar simulation scenario as in \cite{bjornson2019scalable} where $L$ APs and $K$ UEs are independently and uniformly distributed in a $1 \times 1$ km simulation area.
The transmit power for each UE is at most ${p_{\max }} =100$ mW. The channel bandwidth ${{B}} = 20$ MHz.
Each AP is equipped with $N$ antennas.
We assume that the ${\tau _p} < K$ pilot sequences are randomly assigned to the UEs. Inspired by \cite{interdonato2020local}, the UE grouping strategy in FZF is based on the rule
\begin{equation}
\sum\limits_{k = 1}^{\left| {{{\bar {\cal S}}_l}} \right|} {\frac{{{{\bar \beta }_{kl}}}}{{\sum\limits_{t = 1}^K {{\beta _{tl}}} }}}  \ge v\% ,
\end{equation}
where ${\bar {\cal S}}_l$ refers to the set of UEs in ${{\cal S}_l}$ that contribute at least $v$\% of ${\sum\limits_{t = 1}^K {{\beta _{tl}}} }$.
Besides, as mentioned in Remark 3, UEs that use the same pilot are assigned in the same group. We assume $v = 85$.

\begin{figure}[t!]
\centering
\includegraphics[scale=0.6]{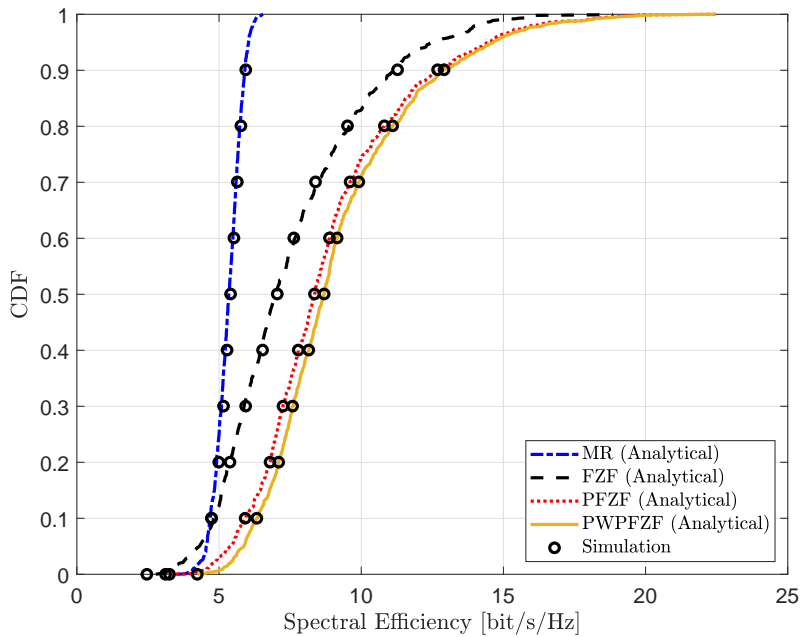}
\caption{CDFs of the SE achieved by the MR, FZF, PFZF and PWPFZF combining schemes with $L=100$, $K=10$, $N=8$, ${\tau_p}=7$ and $p_k =100$ mW for each UE.}
\label{fig1}
\end{figure}

\begin{figure}[t!]
\centering
\includegraphics[scale=0.6]{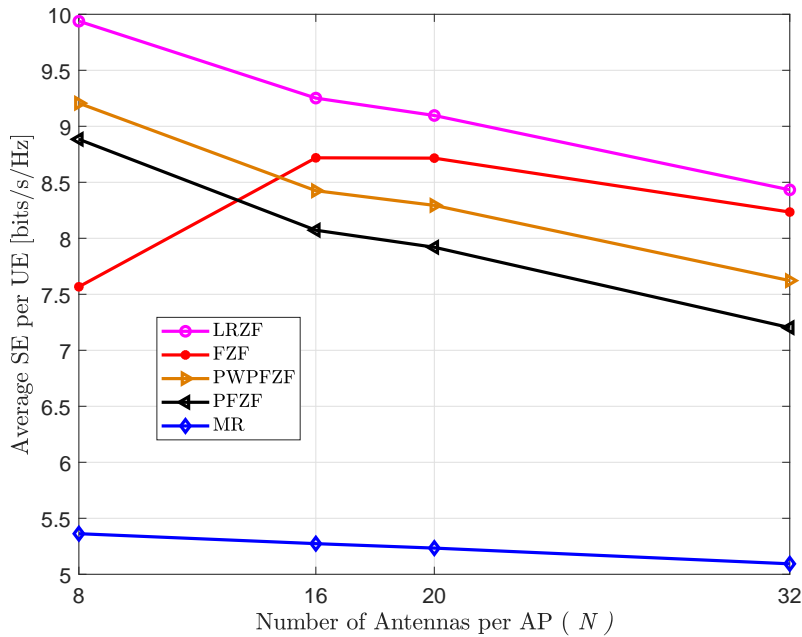}
\caption{Average SE per UE against $N$, with $LN=800$, $K=10$, ${\tau_p}=7$ and $p_k =100$ mW for each UE.}
\label{fig3}
\end{figure}

\begin{figure}[t!]
\centering
\includegraphics[scale=0.6]{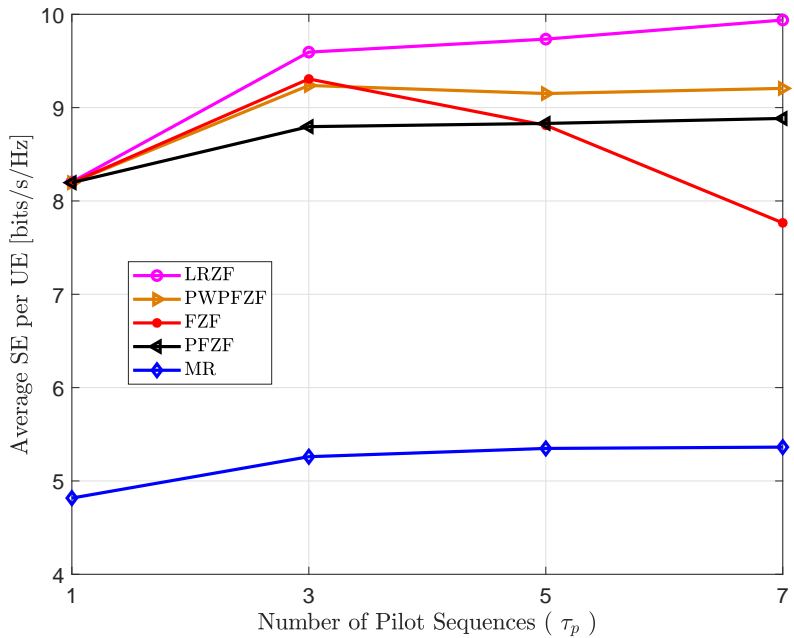}
\caption{Average SE per UE against ${\tau_p}$, with $L=100$, $K=10$, $N=8$ and $p_k =100$ mW for each UE.}
\label{fig4}
\end{figure}

\begin{figure}[t!]
\centering
\includegraphics[scale=0.6]{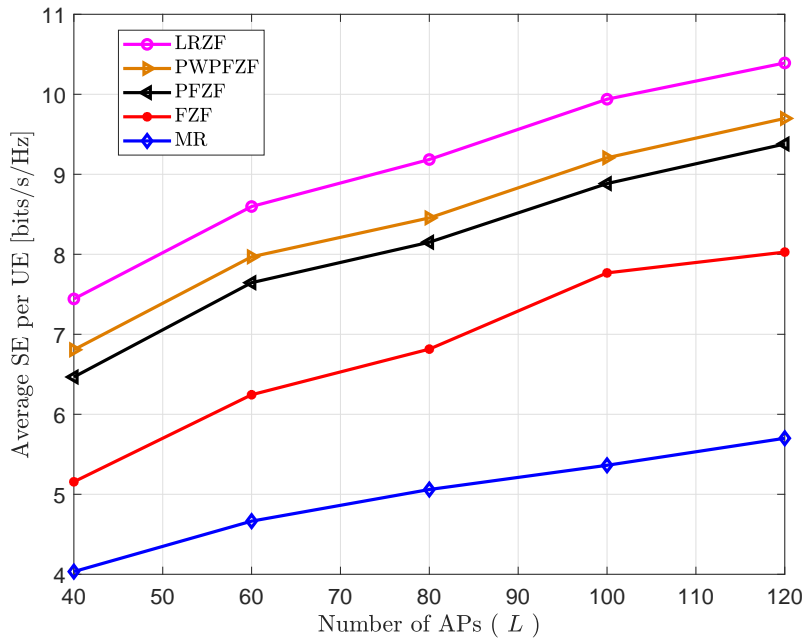}
\caption{Average SE per UE against $L$, with ${\tau_p}=7$, $K=10$, $N=8$ and $p_k =100$ mW for each UE.}
\label{fig5}
\end{figure}

In Fig. \ref{fig1}, the cumulative distribution function (CDF) of the per UE uplink SE is shown for the MR, FZF, PFZF, and PWPFZF combining schemes with $L=100$, $K=10$, $N=8$, ${\tau_p}=7$, and $p_k =100$ mW for each UE.
The performance gap between the MR combining and the ZF-based schemes are quite significant, especially for UEs with large channel gains.
It is resulted from the impact of inter-user interference while FZF, PFZF, and PWPFZF combining schemes all have the ability to suppress that interference.
Specifically, applying FZF combining leads to 41\% improvement in terms of average SE. Besides, the advantage of employing PFZF and PWPFZF rather than FZF is noticeable.
FZF spends ${\tau _p}$ degrees of freedom to cancel the pilot contamination and inter-user interference while PFZF and PWPFZF only spend ${\tau _{{{\cal S}_l}}}$ degrees of freedom and take advantage of a larger array gain.
On the average, the ${\tau _{{{\cal S}_l}}} = 1.5345$.
Compared with PFZF, PWPFZF gives a higher 95\%-likely SE, which is due to its protective nature of weak UEs with lower channel gain.
Furthermore, Fig. \ref{fig1} also shows that the simulations are matching with the numerical results, which proves the accuracy of the Monte Carlo simulations.
Note that the closed-form expression with mLRZF is a large-scale approximation that becomes exact in the asymptotic regime where $N$ and $K$ tend to infinity.
Therefore, LRZF is not considered in Fig. \ref{fig1}.
Furthermore, the results for LRZF are obtained through Monte Carlo simulations in Figs. \ref{fig3}, \ref{fig4}, and \ref{fig5}.

Fig. \ref{fig3} compares the SE of MR, FZF, PFZF, PWPFZF, and LRZF schemes against $N$ with $K=10$, ${\tau_p}=7$, and $p_k =100$ mW for each UE, for systems having the same total numbers of antennas, $LN=800$, but different number of APs.
Focusing on the average SE, we firstly observe that as $N$ increases the performance is in decreasing trend for MR and LRZF combining.
It suggests the high degree of macro-diversity and low path losses are important for offering a high SE.
The performance of PFZF and PWPFZF combining schemes also declines when $N$ increases, which indicates that the macro-diversity gain is dominant over the array gain $N - {\tau _{{{\cal S}_l}}}$.
In contrast, the average SE applying FZF combining with $N=16$ is lower than the case of $N=8$.
This is because that when the number of APs is sufficiently small, the interference is serious and puts back the system performance badly.
When $N$ increases from 8 to 16, the ability to suppress interference also increases, therefore, the average SE is elevated.
It can be seen that the performance gap between FZF and LRZF shrinks when $N$ increases from 8 to 16, which indicates the improvement in suppressing interference.
However, if $N$ still increases, the gain of canceling interference does not dominate the SE.
With the decline of macro-diversity gain, the SE with FZF also decreases. When $N=8, L=100$, PWPFZF performs better than FZF.
Although PWPFZF has a lower average SE than LRZF, it is also a good choice as it has lower computational complexity and can be computed SE expressions in closed-form.
When $N$ increases and $L$ decreases, LRZF is the best choice as it performs better than FZF and has the same computational complexity as FZF.

In Fig. \ref{fig4} we emphasize the impact of the number of pilot sequences on average SE and investigate the average SE against ${\tau_p}$, with $L=100$, $K=10$, $N=8$, and $p_k =100$ mW for each UE.
By increasing ${\tau_p}$, we reduce the pilot re-use, hence the pilot contamination.
Consequently, the ability to suppress the interference increases.
Therefore, the SE with MR and PFZF gain improvement.
Besides, when ${\tau_p} = 1$, the performance of FZF, PFZF, and PWPFZF are identical and very close to LRZF, which is resulted from using the UatF bound.
Along with the increase of ${\tau_p}$, the gain of suppressing the interference helps promote the SE.
However, when ${\tau_p}$ still increases, the array gain $N - {\tau _p}$ reduces in the FZF scheme, which finally leads to a reduction in SE.

Fig. \ref{fig5} shows the average SE with ${\tau_p}=7$, $K=10$, $N=8$ and $p_k =100$ mW for each UE versus $L$, i.e., the number of APs.
For all the combining schemes, it turns out that when increasing the number of APs, the macro-diversity gain increases, and hence, the average SE increases.
As expected, the LRZF provides the highest SE while MR gives the lowest SE.
However, PWPFZF is also a good choice as it has substantially lower computational complexity than LRZF.
Besides, we can compute the SE with PWPFZF in exact closed-form.
In addition, along with the number of APs increases, the FZF, PFZF, PWPFZF, and LRZF gain more in performance than MR since those combining schemes can suppress the interference which becomes more serious as the channel gain of UEs increase.

\begin{figure}[t!]
\centering
\includegraphics[scale=0.6]{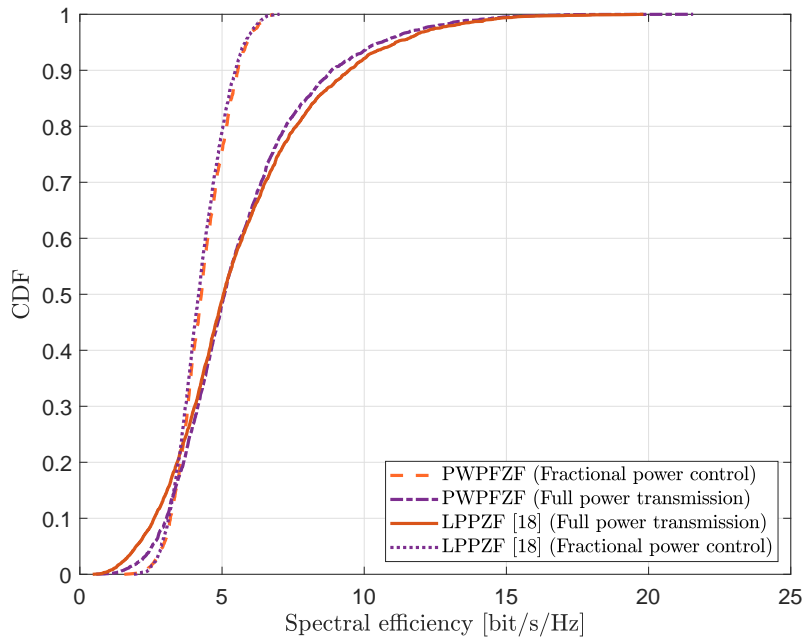}
\caption{CDFs of the SE achieved by the PWPFZF and LPPZF [18] combining scheme with $L = 100$, $K=60$, $N = 8$, ${\tau_p}=7$ and ${p_{\max }} =100$ mW for each UE.}
\label{fig_fractional}
\end{figure}

In Fig. \ref{fig_fractional}, we present the CDF of the per UE uplink SE employed PWPFZF combining scheme with $L = 100$, $K=60$, $N = 8$, ${\tau_p}=7$ and ${p_{\max }} =100$ mW.
The results are achieved by using full power transmission, which means ${p_k} = {p_{\max }}$, $k = 1, \ldots ,K$ and fractional power control which follows the rule of \cite{nikbakht2019uplink}
\begin{equation}
{p_k} \propto \frac{1}{\sum\nolimits_{l = 1}^L {{\beta _{kl}}}},\;\;\;k = 1, \ldots ,K.
\end{equation}
The fractional power control adheres to the rule that the worse the channel is the more power is allocated.
Furthermore, this strategy only depends on the large-scale fading coefficients and is distributed.
In detail, each UE controls its power based on its channel gains and the channel gains of the weakest UE so that the weakest UE uses maximum power.
Therefore, the fractional power control is scalable.
What we observe in the figure is that the SE distribution with fractional power control exceeds the full power transmission at the 95\%-likely SE and the former is much steeper than the latter, which means that fractional power control improves the SE of weak UEs and ensures uniformly good service.
Specifically, with fractional power control, the 95\%-likely SE increases by up to 11.3\% compared with full power transmission.
Besides, although we apply the same UE grouping strategy as \cite{interdonato2020local}, the performance of our PWPFZF and the local partial protective ZF (LPPZF) proposed in \cite{interdonato2020local} is different. The reason is that our PWPFZF protects the weak UEs instead of strong UEs.
Fig. \ref{fig_fractional} shows that when using full power transmission, applying PWPFZF improves the performance of 95\%-likely SE by up to 28\% compared with LPPZF combining.
However, the LPPZF improves the upper SE percentiles compared to PWPFZF, thanks to its protective nature towards the UEs with larger channel gain.

\section{Conclusions}
In this paper, we analyzed the uplink SE of FZF, PFZF, PWPFZF, and LRZF and derived closed-form expression of SE for FZF PFZF, and PWPFZF combining schemes in cell-free massive MIMO under independent Rayleigh channels, addressing channel estimation error and pilot contamination.
The results show that with different configurations, LRZF provides the highest SE while MR gives the lowest SE.
However, when the number of pilot sequences is large and the number of antennas per AP is small, the PWPFZF is also a good choice.
In that case, PWPFZF performs better than MR, FZF, and PFZF.
Although it performs worse than LRZF, it has lower computational complexity and we can compute SE expressions in exact closed-form.
Finally, applying fractional power control can further improve the 95\%-likely SE with PWPFZF combining compared with full power transmission.

\begin{appendices}
\section{}\label{proofFZF}
According to (\ref{FZF}), we first compute the numerator in (\ref{SINR}) in closed-form as
\begin{equation}\label{DS_FZF}
{\left| {{\rm{D}}{{\rm{S}}_k}} \right|^2} = p_k^{{\rm{ul}}}{\left| {\mathbb{E}}{\left\{ {\sum\limits_{l = 1}^L {a_{kl}^*} {{\left( {{\bf{v}}_{{i_k}l}^{{\rm{FZF}}}} \right)}^H}{{\bf{h}}_{kl}}} \right\}} \right|^2} = p_k^{{\rm{ul}}}{\left| {\sum\limits_{l = 1}^L {a_{kl}^*} {{\gamma _{{i_k}l}}}} \right|^2}.\notag
\end{equation}
Then, the variance of the beamforming gain uncertainty is given as
\begin{align}\label{BU_FZF_ori}
{\mathbb{E}}\left\{ {{{\left| {{\rm{B}}{{\rm{U}}_k}} \right|}^2}} \right\}
&=p_k^{{\rm{ul}}}{\mathbb{E}}\left\{ {{{\left| {\sum\limits_{l = 1}^L {a_{kl}^*} {{\left( {{\bf{v}}_{{i_k}l}^{{\rm{FZF}}}} \right)}^H}{{\bf{h}}_{kl}}} \right|}^2}} \right\} - p_k^{{\rm{ul}}}{\left|{\mathbb{E}}{\left\{ {\sum\limits_{l = 1}^L {a_{kl}^*} {{\left( {{\bf{v}}_{{i_k}l}^{{\rm{FZF}}}} \right)}^H}{{\bf{h}}_{kl}}} \right\}} \right|^2}\notag\\
& = p_k^{{\rm{ul}}}{{\cal T}_1} - {\left| {{\rm{D}}{{\rm{S}}_k}} \right|^2},
\end{align}
where
\begin{equation}
{{\cal T}_1} \buildrel \Delta \over = {\mathbb{E}}\left\{ {{{\left| {\sum\limits_{l = 1}^L {a_{kl}^*{{\left( {{\bf{v}}_{{i_k}l}^{{\rm{FZF}}}} \right)}^H}{{\bf{h}}_{kl}}} } \right|}^2}} \right\}.
\end{equation}
We have
\begin{align}\label{T_1_FZF}
{{\cal T}_1} &= {\mathbb{E}}\left\{ {{{\left| {\sum\limits_{l = 1}^L {a_{kl}^*{{\left( {{\bf{v}}_{{i_k}l}^{{\rm{FZF}}}} \right)}^H}\left( {{{{\bf{\hat h}}}_{kl}} + {{{\bf{\tilde h}}}_{kl}}} \right)} } \right|}^2}} \right\}\notag\\
%&= {\mathbb{E}}\left\{ {{{\left| {\sum\limits_{l = 1}^L {a_{kl}^*\left( {{c_{kl}} + {\bf{e}}_{{t_k}}^H{{\left( {{{{\bf{\bar H}}}_l}{\bf{\bar H}}_l^H} \right)}^{ - 1}}{\bf{\bar H}}_l^H{{{\bf{\tilde h}}}_{kl}}} \right)} } \right|}^2}} \right\}\notag\\
&= \sum\limits_{l = 1}^L {\sum\limits_{l' = 1}^L {a_{kl}^*a_{kl'}^*\left( {{\gamma _{{i_k}l}}{\gamma _{{i_k}l'}} + {c_{{i_k}l}}{c_{{i_k}l'}}{\theta _{{i_k}l}}{\theta _{{i_k}l'}}}\right.}}{{\left.{\times\mathbb{E}\left\{ {{\mathbf{\tilde h}}_{kl}^H{{{\mathbf{\bar H}}}_l}{{\left( {{\mathbf{\bar H}}_l^H{{{\mathbf{\bar H}}}_l}} \right)}^{ - 1}}{{\mathbf{e}}_{{i_k}}}{\mathbf{e}}_{{i_k}}^H{{\left( {{{{\mathbf{\bar H}}}_l}{\mathbf{\bar H}}_l^H} \right)}^{ - 1}}{\mathbf{\bar H}}_l^H{{{\mathbf{\tilde h}}}_{kl}}} \right\}} \right)} } \notag\\
&\mathop  = \sum\limits_{l = 1}^L {\sum\limits_{l' = 1}^L {a_{kl}^*a_{kl'}^*\left( {{\gamma _{{i_k}l}}{\gamma _{{i_k}l'}} + \frac{{{c_{{i_k}l}}{c_{{i_k}l'}}{\theta _{kl'}}\left( {{\beta _{kl}} - {\gamma _{kl}}} \right)}}{{\left( {N - {\tau _p}} \right)}}} \right)} } \notag\\
&={\left( {\sum\limits_{l = 1}^L {a_{kl}^*{\gamma _{{i_k}l}}} } \right)^2} + \sum\limits_{l = 1}^L {{{\left| {a_{kl}^*} \right|}^2}} \frac{{{\gamma _{{i_k}l}}\left( {{\beta _{kl}} - {\gamma _{{i_k}l}}} \right)}}{{\left( {N - {\tau _p}} \right)}},
\end{align}
where ${\left( a \right)}$ is computed by applying
\begin{equation}
{\mathbb{E}}\left\{ {{{\left\| {{{{\bf{\bar H}}}_l}{{\left( {{\bf{\bar H}}_l^H{{{\bf{\bar H}}}_l}} \right)}^{ - 1}}{{\bf{e}}_{{i_k}}}} \right\|}^2}} \right\}{\rm{ = }}\frac{1}{{\left( {N - {\tau _p}} \right){\theta _{kl}}}},
\end{equation}
which follows from \cite[Lemma 2.10]{Tulino2004random}, for a ${\tau _p} \times {\tau _p}$ central complex Wishart matrix with $N$ degrees of freedom satisfying $N \ge {\tau _p} + 1$. Substitution (\ref{T_1_FZF}) and (\ref{DS_FZF}) into (\ref{BU_FZF_ori}) yields
\begin{equation}
{\mathbb{E}}\left\{ {{{\left| {{\rm{B}}{{\rm{U}}_k}} \right|}^2}} \right\} = p_k^{{\rm{ul}}}\sum\limits_{l = 1}^L {{{\left| {a_{kl}^*} \right|}^2}} \frac{{{\gamma _{kl}}\left( {{\beta _{kl}} - {\gamma _{kl}}} \right)}}{{\left( {N - {\tau _p}} \right)}}.
\end{equation}
Next, we compute the variance of pilot contamination term as
\begin{align}
\sum\limits_{t \in {{\cal P}_k}/\left\{ k \right\}}^K {\mathbb{E}}{\left\{ {{{\left| {{\rm{P}}{{\rm{C}}_k}} \right|}^2}} \right\}}  &= \sum\limits_{t \in {{\cal P}_k}/\left\{ k \right\}}^K {p_t^{{\rm{ul}}}{\mathbb{E}}\left\{ {{{\left| {\sum\limits_{l = 1}^L {a_{kl}^*} {{\left( {{\bf{v}}_{{i_k}l}^{{\rm{FZF}}}} \right)}^H}{{\bf{h}}_{tl}}} \right|}^2}} \right\}}\notag\\
& = \sum\limits_{t \in {{\cal P}_k}/\left\{ k \right\}}^K {p_t^{{\rm{ul}}}{\mathbb{E}}\left\{ {{{\left| {\sum\limits_{l = 1}^L {a_{kl}^*} {{\left( {{\bf{v}}_{{i_k}l}^{{\rm{FZF}}}} \right)}^H}\left( {{{{\bf{\hat h}}}_{tl}} + {{{\bf{\tilde h}}}_{tl}}} \right)} \right|}^2}} \right\}} \notag\\
& = \sum\limits_{t \in {\mathcal{P}_k}/\left\{ k \right\}}^K {p_i^{{\rm{ul}}}{{\left( {\sum\limits_{l = 1}^L {a_{kl}^*{\gamma _{tl}}} } \right)}^2}}  + \sum\limits_{t \in {\mathcal{P}_k}/\left\{ k \right\}}^K {p_i^{{\rm{ul}}}\sum\limits_{l = 1}^L {{{\left| {a_{kl}^*} \right|}^2}} \frac{{{\gamma _{kl}}\left( {{\beta _{tl}} - {\gamma _{tl}}} \right)}}{{\left( {N - {\tau _p}} \right)}}}.
\end{align}
Similarly, the interference caused by UEs that use different pilots can be computed as
\begin{align}
\sum\limits_{t \notin {{\cal P}_k}}^K {\mathbb{E}}{\left\{ {{{\left| {{\rm{U}}{{\rm{I}}_k}} \right|}^2}} \right\}} &= \sum\limits_{t \notin {{\cal P}_k}}^K {p_t^{{\rm{ul}}}{\mathbb{E}}\left\{ {{{\left| {\sum\limits_{l = 1}^L {a_{kl}^*} {{\left( {{\bf{v}}_{{i_k}l}^{{\rm{FZF}}}} \right)}^H}{{\bf{h}}_{tl}}} \right|}^2}} \right\}}= \sum\limits_{t \notin {\mathcal{P}_k}}^K {p_i^{{\rm{ul}}}\sum\limits_{l = 1}^L {{{\left| {a_{kl}^*} \right|}^2}} \frac{{{\gamma _{kl}}\left( {{\beta _{tl}} - {\gamma _{tl}}} \right)}}{{\left( {N - {\tau _p}} \right)}}} .
\end{align}
The last expectation in the denominator of (\ref{SINR}) is computed using the fact that the noise and the channel estimate are independent, leading to
\begin{align}
{\mathbb{E}}\left\{ {{{\left| {{\rm{G}}{{\rm{N}}_k}} \right|}^2}} \right\} &= {\mathbb{E}}\left\{ {{{\left| {\sum\limits_{l = 1}^L {a_{kl}^*} {{\left( {{\bf{v}}_{{i_k}l}^{{\rm{FZF}}}} \right)}^H}{{\bf{n}}_l}} \right|}^2}} \right\} = {\sigma ^2}\sum\limits_{l = 1}^L {{{\left| {a_{kl}^*} \right|}^2}\frac{{{\gamma _{kl}}}}{{\left( {N - {\tau _p}} \right)}}} .
\end{align}
Finally, we can rewrite (\ref{SINR}) as (\ref{SINR_FZF}). After simplification, (\ref{SINR_FZF}) can be expressed as ${p_k^{{\rm{ul}}}{\bf{b}}_k^H{\bf{C}}_k^{ - 1}{{\bf{b}}_k}}$. The proof is concluded by deriving the closed-form SE expression in (\ref{SINR_max}).

\section{}\label{proofPFZF}
Plugging (\ref{V_PFZF_S}) and (\ref{V_PFZF_W}) into (\ref{E}) yields
\begin{align}\label{DS_PFZF}
{\left| {{\rm{D}}{{\rm{S}}_k}} \right|^2}
&= p_k^{{\rm{ul}}}{\left| {\mathbb{E}}{\left\{ {\sum\limits_{l \in {{\cal Z}_k}} {a_{kl}^*} {{{\left( {{\bf{v}}_{{i_k}l}^{{\rm{FZF}}}} \right)}^H}}{{\bf{h}}_{kl}} + \sum\limits_{l \in {{\cal M}_k}} {a_{kl}^*} {{\left( {{\bf{v}}_{kl}^{{\rm{MR}}}} \right)}^H}{{\bf{h}}_{kl}}} \right\}} \right|^2} \notag\\
&= p_k^{{\rm{ul}}}{\left| {\sum\limits_{l \in {\mathcal{Z}_k}} {a_{kl}^*{\gamma _{kl}}}  + \sum\limits_{l \in {\mathcal{M}_k}} {Na_{kl}^*{\gamma _{kl}}} } \right|^2}.
\end{align}
The first term of the denominator in (\ref{SINR}) can be computed as
\begin{align}\label{BU_PFZF_ori}
{\mathbb{E}}\left\{ {{{\left| {{\rm{B}}{{\rm{U}}_k}} \right|}^2}} \right\}
%&= p_k^{{\rm{ul}}}{\mathbb{E}}\left\{ {\left| {\left( {\sum\limits_{l \in {{\cal Z}_k}} {a_{kl}^*} {{\left( {{\bf{v}}_{{i_k}l}^{{\rm{PFZF}}}} \right)}^H}{{\bf{h}}_{kl}} + \sum\limits_{l \in {{\cal M}_k}} {a_{kl}^*} {{\left( {{\bf{v}}_{kl}^{{\rm{MR}}}} \right)}^H}{{\bf{h}}_{kl}}} \right)} \right.} \right.\notag\\
%& \left. { - {\mathbb{E}}{{\left. {\left\{ {\sum\limits_{l \in {{\cal Z}_k}} {a_{kl}^*} {{\left( {{\bf{v}}_{{i_k}l}^{{\rm{PFZF}}}} \right)}^H}{{\bf{h}}_{kl}} + \sum\limits_{l \in {{\cal M}_k}} {a_{kl}^*} {{\left( {{\bf{v}}_{kl}^{{\rm{MR}}}} \right)}^H}{{\bf{h}}_{kl}}} \right\}} \right|}^2}} \right\}\notag\\
&=p_k^{{\rm{ul}}}{\mathbb{E}}\left\{ {{{\left| {\sum\limits_{l \in {{\cal Z}_k}} {a_{kl}^*} {{\left( {{\bf{v}}_{{i_k}l}^{{\rm{PFZF}}}} \right)}^H}{{\bf{h}}_{kl}} + \sum\limits_{l \in {{\cal M}_k}} {a_{kl}^*} {{\left( {{\bf{v}}_{kl}^{{\rm{MR}}}} \right)}^H}{{\bf{h}}_{kl}}} \right|}^2}} \right\} \notag\\
&- p_k^{{\rm{ul}}}{\left| {\mathbb{E}}{\left\{ {\sum\limits_{l \in {{\cal Z}_k}} {a_{kl}^*} {{\left(  {{\bf{v}}_{{i_k}l}^{{\rm{PFZF}}}} \right)}^H}{{\bf{h}}_{kl}} + \sum\limits_{l \in {{\cal M}_k}} {a_{kl}^*} {{\left(  {{\bf{v}}_{kl}^{{\rm{MR}}}} \right)}^H}{{\bf{h}}_{kl}}} \right\}} \right|^2}.
\end{align}
We have
\begin{align}\label{BU_PFZF_2}
&{\mathbb{E}}\left\{ {{{\left| {\sum\limits_{l \in {{\cal Z}_k}} {a_{kl}^*} {{\left( {{\bf{v}}_{{i_k}l}^{{\rm{PFZF}}}} \right)}^H}{{\bf{h}}_{kl}} + \sum\limits_{l \in {{\cal M}_k}} {a_{kl}^*} {{\left( {{\bf{v}}_{kl}^{{\rm{MR}}}} \right)}^H}{{\bf{h}}_{kl}}} \right|}^2}} \right\}\notag\\
&\mathop  = \limits^{\left( b \right)} p_k^{{\rm{ul}}}\left( {{{\left( {\sum\limits_{l \in {{\cal Z}_k}} {a_{kl}^*{c_{kl}}} } \right)}^2}+ \sum\limits_{l \in {{\cal Z}_k}} {{{\left| {a_{kl}^*} \right|}^2}} \frac{{{\beta _{kl}} - {\gamma _{kl}}}}{{\left( {N - {{\tau _{{\mathcal{S}_l}}}}} \right){\theta _{kl}}}}} \right) \notag\\
&+ p_k^{{\rm{ul}}}\left( {\sum\limits_{l \in {\mathcal{M}_k}} {{{\left| {a_{kl}^*} \right|}^2}N{\beta _{kl}}{\gamma _{kl}}}  +  {{\left| {\sum\limits_{l \in {\mathcal{M}_k}} {a_{kl}^*} \mathbb{E}\left\{ {{{\left( {{\mathbf{v}}_{kl}^{{\rm{MR}}}} \right)}^H}{{\mathbf{h}}_{kl}}} \right\}} \right|}^2}} \right)\notag\\
&+ 2\left( {\sum\limits_{l \in {{\cal Z}_k}} {a_{kl}^*} {c_{kl}}} \right)\left( {\sum\limits_{l \in {{\cal M}_k}} {a_{kl}^*N{\gamma _{kl}}} } \right),
\end{align}
where ${\left( b \right)}$ follows the fact
\begin{equation}
{\mathbb{E}}\left\{ {{{\left\| {{{{\bf{\bar H}}}_l}{{\bf{E}}_{{{\cal S}_l}}}{{\left( {{\bf{E}}_{{{\cal S}_l}}^H{{{\bf{\bar H}}}_l}{{{\bf{\bar H}}}_l}{{\bf{E}}_{{{\cal S}_l}}}} \right)}^{ - 1}}{\varepsilon _{{j_{kl}}}}} \right\|}^2}} \right\} =  \frac{1}{{\left( {N - {\tau _{{{\cal S}_l}}}} \right){\theta _{kl}}}},
\end{equation}
which follows from \cite[Lemma 2.10]{Tulino2004random}, for a ${\tau _{{{\cal S}_l}}} \times {\tau _{{{\cal S}_l}}}$ central complex Wishart matrix with $N$ degrees of freedom satisfying $N \ge {\tau _{{{\cal S}_l}}} + 1$, and ${\mathbb{E}}\left\{ {{{\left| {\sum\nolimits_{l \in {{\cal M}_k}} {a_{kl}^*} {{\left( {{\bf{v}}_{kl}^{{\rm{MR}}}} \right)}^H}{{\bf{h}}_{kl}}} \right|}^2}} \right\}$ is computed as
\begin{align}
&{\mathbb{E}}\left\{ {{{\left| {\sum\limits_{l \in {{\cal M}_k}} {a_{kl}^*} {{\left( {{\bf{v}}_{kl}^{{\rm{MR}}}} \right)}^H}{{\bf{h}}_{kl}}} \right|}^2}} \right\} \notag\\
&= \sum\limits_{l \in {{\cal M}_k}} {{{\left| {a_{kl}^*} \right|}^2}{\mathbb{E}}\left\{ {{{\left| {{\bf{\hat h}}_{kl}^H{{\bf{h}}_{kl}}} \right|}^2}} \right\}}  + {\left| {\sum\limits_{l \in {{\cal Z}_k}} {a_{kl}^*{\mathbb{E}}\left\{ {{\bf{\hat h}}_{kl}^H{{\bf{h}}_{kl}}} \right\}} } \right|^2}- \sum\limits_{l \in {{\cal M}_k}} {{{\left| {a_{kl}^*{\mathbb{E}}\left\{ {{\bf{\hat h}}_{kl}^H{{\bf{h}}_{kl}}} \right\}} \right|}^2}}\notag\\
&\mathop  = \limits^{\left( c \right)} \sum\limits_{l \in {{\cal M}_k}} {{{\left| {a_{kl}^*} \right|}^2}N{\beta _{kl}}{\gamma _{kl}}}  + {\left| {\sum\limits_{l \in {{\cal M}_k}} {Na_{kl}^*} {\gamma _{kl}}} \right|^2},
\end{align}
where ${\left( c \right)}$ is computed by applying the property in \cite[Appendix A]{ozdogan2019massive}.
Substitution (\ref{DS_PFZF}) and (\ref{BU_PFZF_2}) into (\ref{BU_PFZF_ori}) yields
\begin{align}\label{BU_PFZF}
&{\mathbb{E}}\left\{ {{{\left| {{\rm{B}}{{\rm{U}}_k}} \right|}^2}} \right\} = p_k^{{\rm{ul}}}\sum\limits_{l \in {\mathcal{Z}_k}} {{{\left| {a_{kl}^*} \right|}^2}} \frac{{{\gamma _{kl}}\left( {{\beta _{kl}} - {\gamma _{kl}}} \right)}}{{\left( {N - {\tau _p}} \right)}} + p_k^{{\rm{ul}}}\sum\limits_{l \in {\mathcal{M}_k}} {{{\left| {a_{kl}^*} \right|}^2}N{\beta _{kl}}{\gamma _{kl}}} .
\end{align}
The variance of the pilot contamination in the denominator of (\ref{SINR}) is computed as
\begin{align}\label{PC_PFZF_ori}
\sum\limits_{t \in {{\cal P}_k}/\left\{ k \right\}}^K {\mathbb{E}}{\left\{ {{{\left| {{\rm{P}}{{\rm{C}}_k}} \right|}^2}} \right\}}&= \sum\limits_{t \in {{\cal P}_k}/\left\{ k \right\}}^K {p_t^{{\rm{ul}}}{\mathbb{E}}\left\{ {{{\left| {\sum\limits_{l \in {{\cal Z}_k}} {a_{kl}^*} {{\left( {{\bf{v}}_{{i_k}l}^{{\rm{PFZF}}}} \right)}^H}{{\bf{h}}_{tl}} + \sum\limits_{l \in {{\cal M}_k}} {a_{kl}^*} {{\left( {{\bf{v}}_{kl}^{{\rm{MR}}}} \right)}^H}{{\bf{h}}_{tl}}} \right|}^2}} \right\}} \notag\\
&= \sum\limits_{t \in {{\cal P}_k}/\left\{ k \right\}}^K {p_t^{{\rm{ul}}}{\mathbb{E}}\left\{ {{{\left| {\sum\limits_{l \in {{\cal Z}_k}} {a_{kl}^*} {{\left( {{\bf{v}}_{{i_k}l}^{{\rm{PFZF}}}} \right)}^H}{{\bf{h}}_{tl}}} \right|}^2}} \right\}}  + {{\cal T}_2} +  {{\cal T}_3},
\end{align}
where
\begin{align}
&{{\cal T}_2} = \sum\limits_{t \in {{\cal P}_k}/\left\{ k \right\}}^K {p_t^{{\rm{ul}}}{\mathbb{E}}\left\{ {{{\left| {\sum\limits_{l \in {{\cal M}_k}} {a_{kl}^*} {{\left( {{\bf{v}}_{kl}^{{\rm{MR}}}} \right)}^H}{{\bf{h}}_{tl}}} \right|}^2}} \right\}},\notag\\
&{{\cal T}_3} = 2\sum\limits_{t \in {{\cal P}_k}/\left\{ k \right\}}^K {p_t^{{\rm{ul}}}{\mathbb{E}}\left\{ {\left( {\sum\limits_{l \in {{\cal Z}_k}} {a_{kl}^*} {{\left(  {{\bf{v}}_{{i_k}l}^{{\rm{PFZF}}}} \right)}^H}{{\bf{h}}_{tl}}} \right)\left( {\sum\limits_{l \in {{\cal M}_k}} {a_{kl}^*} {{\left( {{\bf{v}}_{kl}^{{\rm{MR}}}} \right)}^H}{{\bf{h}}_{tl}}} \right)} \right\}}.
\end{align}
The expectation ${{\cal T}_2}$ and ${{\cal T}_3}$ is computed by applying the property in \cite[Appendix A]{ozdogan2019massive} and
\begin{align}
&{\mathbb{E}}\left\{ {{{\left( {{\bf{v}}_{kl}^{{\rm{MR}}}} \right)}^H}{{\bf{h}}_{tl}}} \right\} = {\mathbb{E}}\left\{ {{\bf{\hat h}}_{kl}^H{{{\bf{\hat h}}}_{tl}}} \right\} = \frac{{{c_{kl}}{c_{tl}}}}{{{\tau _p}}}{\mathbb{E}}\left\{ {{{\left( {{{\bf{Z}}_l}{{\bm{\phi}} _{{i_k}}}} \right)}^H}\left( {{{\bf{Z}}_l}{{\bm{\phi}} _{{i_k}}}} \right)} \right\} =N{\gamma _{kl}},
\end{align}
as
\begin{align}\label{T2_PFZF}
&{{\cal T}_2}  = \sum\limits_{l \in {\mathcal{M}_k}} {{{\left| {a_{kl}^*} \right|}^2}} \left( {{N^2}\gamma _{{i_k}}^2 + N{\gamma _{kl}}{\beta _{tl}}} \right) + {\left| {\sum\limits_{l \in {\mathcal{M}_k}} {a_{kl}^*} N{\gamma _{kl}}} \right|^2} - \sum\limits_{l \in {\mathcal{M}_k}} {{{\left| {a_{kl}^*N{\gamma _{{i_k}l}}} \right|}^2}},
\end{align}
and
\begin{equation}\label{T3_PFZF}
{{\cal T}_3} = {\rm{2}}\sum\limits_{t \in {{\cal P}_k}/\left\{ k \right\}}^K {p_t^{{\rm{ul}}}}N\left( {\sum\limits_{l \in {\mathcal{Z}_k}} {a_{kl}^*{\gamma _{tl}}} } \right)\left( {\sum\limits_{l \in {\mathcal{M}_k}} {a_{kl}^*{\gamma _{tl}}} } \right).
\end{equation}
Substitution (\ref{T2_PFZF}) and (\ref{T3_PFZF}) into (\ref{PC_PFZF_ori}) yields
\begin{align}\label{PC_PFZF}
\sum\limits_{t \in {{\cal P}_k}/\left\{ k \right\}}^K {\mathbb{E}}{\left\{ {{{\left| {{\rm{P}}{{\rm{C}}_k}} \right|}^2}} \right\}}
&= \sum\limits_{t \in {\mathcal{P}_k}/\left\{ k \right\}}^K {p_i^{{\rm{ul}}}} \left( {{{\left( {\sum\limits_{l \in {\mathcal{Z}_k}} {a_{kl}^*{\gamma _{tl}}}  + N\sum\limits_{l \in {\mathcal{M}_k}} {a_{kl}^*{\gamma _{tl}}} } \right)}^2}} \right.\notag\\
&{\times}\left. {\sum\limits_{l \in {\mathcal{Z}_k}} {{{\left| {a_{kl}^*} \right|}^2}\frac{{{\gamma _{kl}}\left( {{\beta _{tl}} - {\gamma _{tl}}} \right)}}{{\left( {N - {\tau _{{\mathcal{S}_l}}}} \right)}}}  + \sum\limits_{l \in {\mathcal{M}_k}} {{{\left| {a_{kl}^*} \right|}^2}} N{\gamma _{kl}}{\beta _{tl}}} \right).
\end{align}
Similarly, the inter-user interference caused by other UEs that use different pilots can be computed as
\begin{align}\label{MU_PFZF}
&\sum\limits_{t \notin {{\cal P}_k}}^K {\mathbb{E}}{\left\{ {{{\left| {{\rm{U}}{{\rm{I}}_k}} \right|}^2}} \right\}}
%&= \sum\limits_{t \notin {{\cal P}_k}}^K {p_i^{{\rm{ul}}}} {\mathbb{E}}\left\{ {{{\left| {\sum\limits_{l \in {{\cal Z}_k}} {a_{kl}^*} {{\left( {{\bf{v}}_{{i_k}l}^{{\rm{PFZF}}}} \right)}^H}{{\bf{h}}_{tl}} +  \sum\limits_{l \in {{\cal W}_k}} {a_{kl}^*} {{\left( {{\bf{v}}_{kl}^{{\rm{MR}}}} \right)}^H}{{\bf{h}}_{tl}}} \right|}^2}} \right\}\notag\\
= \sum\limits_{t \notin {{\cal P}_k}}^K {p_t^{{\rm{ul}}}} \left( {\sum\limits_{l \in {{\cal Z}_k}} {{{\left| {a_{kl}^*} \right|}^2}} {\frac{{{\gamma _{kl}}\left( {{\beta _{tl}} - {\gamma _{tl}}} \right)}}{{\left( {N - {\tau _{{\mathcal{S}_L}}}} \right)}}} +  \sum\limits_{l \in {{\cal W}_k}} {{{\left| {a_{kl}^*} \right|}^2}} N{\gamma _{kl}}{\beta _{tl}}} \right).
\end{align}
The last expectation in the denominator is computed as
\begin{align}\label{GN_PFZF}
{\mathbb{E}}\left\{ {{{\left| {{\rm{G}}{{\rm{N}}_k}} \right|}^2}} \right\}
%&= {\mathbb{E}}\left\{ {{{\left| {\left( {\sum\limits_{l \in {{\cal Z}_k}} {a_{kl}^*} {{\left( {{\bf{v}}_{{i_k}l}^{{\rm{PFZF}}}} \right)}^H}{{\bf{n}}_l} + \sum\limits_{l \in {{\cal M}_k}} {a_{kl}^*} {{\left( {{\bf{v}}_{kl}^{{\rm{MR}}}} \right)}^H}{{\bf{n}}_l}} \right)} \right|}^2}} \right\}\notag\\
&= {\mathbb{E}}\left\{ {{{\left| {\sum\limits_{l \in {{\cal Z}_k}} {a_{kl}^*} {{\left( {{\bf{v}}_{{i_k}l}^{{\rm{PFZF}}}} \right)}^H}{{\bf{n}}_l}} \right|}^2}} \right\} + {\mathbb{E}}\left\{ {{{\left| {\sum\limits_{l \in {{\cal M}_k}} {a_{kl}^*} {{\left( {{\bf{v}}_{kl}^{{\rm{MR}}}} \right)}^H}{{\bf{n}}_l}} \right|}^2}} \right\}\notag\\
& = {\sigma ^2}\sum\limits_{l \in {\mathcal{Z}_k}} {\frac{{{{\left| {a_{kl}^*} \right|}^2}{\gamma _{kl}}}}{{\left( {N - {\tau _{{\mathcal{S}_l}}}} \right)}}}  + {\sigma ^2}\sum\limits_{l \in {\mathcal{M}_k}} {{{\left| {a_{kl}^*} \right|}^2}} N{\gamma _{kl}}.
\end{align}
Using (\ref{DS_PFZF}), (\ref{BU_PFZF}), (\ref{PC_PFZF}), (\ref{MU_PFZF}) and (\ref{GN_PFZF}), we can rewrite (\ref{SINR}) as (\ref{SINR_PFZF}) and obtain the closed-form expression for (\ref{SINR_max}) as shown in Corollary 3.

\section{}\label{proofPWPFZF}
The proof of Corollary 4 is almost identical to what is given in Appendix B for Corollary 3.
The only difference is that the combining vector for the MR at AP $l$ now projects the signal to the ${N - {\tau _{{{\cal S}_l}}}}$ dimensional subspace orthogonal to the column space of ${{{\bf{\bar H}}}_l}{{\bf{E}}_{{{\cal S}_l}}}$.
This projection implies that, for any UE $t$, $k \in {{\cal W}_l}$, with $t \in {{\cal P}_k}$, we have
\begin{align}
&{\mathbb{E}}\left\{ {{{\left( {{\bf{v}}_{kl}^{{\rm{PMRT}}}} \right)}^H}{{{\bf{\hat h}}}_{tl}}} \right\} = {\gamma _{kl}}\left( {N - {\tau _{{{\cal S}_l}}}} \right), \notag\\
&{\mathbb{E}}\left\{ {{{\left| {{{\left( {{\bf{v}}_{kl}^{{\rm{PMRT}}}} \right)}^H}{{{\bf{\hat h}}}_{tl}}} \right|}^2}} \right\}\mathop  = \limits^{\left( d \right)} \left( {N - {\tau _{{{\cal S}_l}}}} \right)\left( {N - {\tau _{{{\cal S}_l}}} + 1} \right)\gamma _{kl}^2,
\end{align}
where $\left( d \right)$ follows from \cite[Lemma 2.9]{Tulino2004random}, for a ${\tau _{{{\cal S}_l}}} \times {\tau _{{{\cal S}_l}}}$ central complex Wishart matrix with $N$ degrees of freedom satisfying $N \ge {\tau _{{{\cal S}_l}}} + 1$. If $k \in {{\cal W}_l}$ and $t \in {{\cal S}_l}$, , with $t  \notin  {{\cal P}_k}$, we have
\begin{align}
{\mathbb{E}}\left\{ {{{\left| {{{\left( {{\bf{v}}_{kl}^{{\rm{PMRT}}}} \right)}^H}{{\bf{h}}_{tl}}} \right|}^2}} \right\}
&= {\mathbb{E}}\left\{ {{{\left| {{{\left( {{\bf{v}}_{kl}^{{\rm{PMRT}}}} \right)}^H}{{{\bf{\tilde h}}}_{tl}}} \right|}^2}} \right\} = \left( {N - {\tau _{{{\cal S}_l}}}} \right)\left( {{\beta _{tl}} - {\gamma _{tl}}} \right){\gamma _{kl}},
\end{align}
since, by design, ${{{\left( {{\bf{v}}_{kl}^{{\rm{PMRT}}}} \right)}^H}{{{\bf{\hat h}}}_{tl}}}=0$, and ${{{\bf{\tilde h}}}_{tl}}$ is independent of ${{\bf{v}}_{kl}^{{\rm{PMRT}}}}$. Other calculations are identical to Appendix B.
\end{appendices}


\begin{thebibliography}{1}

\bibitem{zhang2020Prospective}
J.~{Zhang}, E.~{Bj{\"o}rnson}, M.~{Matthaiou}, D.~W.~K. {Ng}, H.~{Yang}, and
  D.~J. {Love}, ``Prospective multiple antenna technologies for beyond {5G},''
  \emph{IEEE J. Sel. Areas Commun.}, vol.~38, no.~8, pp. 1637--1660, Aug. 2020.

\bibitem{Lopez-Perez2011Enhanced}
{D. Lopez-Perez, I. Guvenc, G. de la Roche, M. Kountouris, T. Quek, and J.
  Zhang}, ``Enhanced intercell interference coordination challenges in
  heterogeneous networks,'' \emph{IEEE Wireless Commun.}, vol.~18, no.~3, pp.
  22--30, Jun. 2011.

\bibitem{ngo2017on}
H.~Q. Ngo, L.~Tran, T.~Q. Duong, M.~Matthaiou, and E.~G. Larsson, ``On the
  total energy efficiency of cell-free massive {MIMO},'' \emph{IEEE Trans.
  Green Commum. Netw.}, vol.~2, no.~1, pp. 25--39, Mar. 2018.

\bibitem{Andrews2016Are}
{J. G. Andrews, X. Zhang, G. D. Durgin, and A. K. Gupta}, ``Are we approaching
  the fundamental limits of wireless network densification?'' \emph{IEEE
  Commun. Mag.}, vol.~54, no.~10, pp. 184--190, Oct. 2016.

\bibitem{mai2020downlink}
T.~C. Mai, H.~Q. Ngo, and T.~Q. Duong, ``Downlink spectral efficiency of
  cell-free massive {MIMO} systems with multi-antenna users,'' \emph{IEEE
  Trans. Commun.}, vol.~68, no.~8, pp. 4803--4815, Aug. 2020.

\bibitem{venkatesan2007network}
S.~Venkatesan, A.~Lozano, and R.~A. Valenzuela, ``Network {MIMO}: Overcoming
  intercell interference in indoor wireless systems,'' \emph{Proc. IEEE ACSSC},
  pp. 83--87, 2007.

\bibitem{nayebi2017precoding}
E.~Nayebi, A.~Ashikhmin, T.~L. Marzetta, H.~Yang, and B.~D. Rao, ``Precoding
  and power optimization in cell-free massive {MIMO} systems,'' \emph{IEEE
  Trans. Wireless Commun.}, vol.~16, no.~7, pp. 4445--4459, Jul. 2017.

\bibitem{bjornson2019making}
E.~Bj{\"o}rnson and L.~Sanguinetti, ``Making cell-free massive {MIMO}
  competitive with {MMSE} processing and centralized implementation,''
  \emph{IEEE Trans. Wireless Commun.}, vol.~19, no.~1, pp. 77--90, Jan. 2020.

\bibitem{gesbert2010multi-cell}
D.~Gesbert, S.~V. Hanly, H.~Huang, S.~S. Shitz, O.~Simeone, and W.~Yu,
  ``Multi-cell {MIMO} cooperative networks: A new look at interference,''
  \emph{IEEE J. Sel. Areas Commun.}, vol.~28, no.~9, pp. 1380--1408, Sep. 2010.

\bibitem{bjornson2019scalable}
E.~Bj{\"o}rnson and L.~Sanguinetti, ``Scalable cell-free massive {MIMO}
  systems,'' \emph{IEEE Trans. Commun.}, vol.~68, no.~7, pp. 4247--4261, Jul.
  2020.

\bibitem{ngo2017cell}
H.~Q. Ngo, A.~Ashikhmin, H.~Yang, E.~G. Larsson, and T.~L. Marzetta,
  ``Cell-free massive {MIMO} versus small cells,'' \emph{IEEE Trans. Wireless
  Commun.}, vol.~16, no.~3, pp. 1834--1850, Mar. 2017.

\bibitem{papazafeiropoulos2020towards}
A.~Papazafeiropoulos, H.~Q. Ngo, P.~Kourtessis, S.~Chatzinotas, and J.~M.
  Senior, ``Towards optimal energy efficiency in cell-free massive {MIMO}
  systems,'' \emph{Proc. IEEE PIMRC}, Aug. 2020.

\bibitem{Zhang2019cell}
J.~{Zhang}, S.~{Chen}, Y.~{Lin}, J.~{Zheng}, B.~{Ai}, and L.~{Hanzo},
  ``Cell-free massive {MIMO}: {A} new next-generation paradigm,'' \emph{IEEE
  Access}, vol.~7, pp. 99\,878--99\,888, 2019.

\bibitem{Buzzi2018User}
{S. Buzzi, C. D'Andrea, and C. D'Elia}, ``User-centric cell-free massive {MIMO}
  with interference cancellation and local {ZF} downlink precoding,''
  \emph{Proc. IEEE ISWCS}, pp. 1--5, Aug. 2018.

\bibitem{[C1]}
D.~{Maryopi}, M.~{Bashar}, and A.~{Burr}, ``On the uplink throughput of zero
  forcing in cell-free massive {MIMO} with coarse quantization,'' \emph{IEEE
  Trans. Veh. Technol.}, vol.~68, no.~7, pp. 7220--7224, Jul. 2019.

\bibitem{[C2]}
P.~{Liu}, K.~{Luo}, D.~{Chen}, and T.~{Jiang}, ``Spectral efficiency analysis
  of cell-free massive {MIMO} systems with zero-forcing detector,'' \emph{IEEE
  Trans. Wireless Commun.}, vol.~19, no.~2, pp. 795--807, Feb. 2020.

\bibitem{[C4]}
H.~V. {Nguyen}, V.~D. {Nguyen}, O.~A. {Dobre}, S.~K. {Sharma},
  S.~{Chatzinotas}, B.~{Ottersten}, and O.~S. {Shin}, ``On the spectral and
  energy efficiencies of full-duplex cell-free massive {MIMO},'' \emph{IEEE J.
  Sel. Areas Commun.}, vol.~38, no.~8, pp. 1698--1718, Aug. 2020.

\bibitem{interdonato2020local}
G.~Interdonato, M.~Karlsson, E.~Bj{\"o}rnson, and E.~G. Larsson, ``Local
  partial zero-forcing precoding for cell-free massive {MIMO},'' \emph{IEEE
  Trans. Wireless Commun.}, vol.~19, no.~7, pp. 4758--4774, Jul. 2020.

\bibitem{[C3]}
F.~{Rezaei}, C.~{Tellambura}, A.~A. {Tadaion}, and A.~R. {Heidarpour}, ``Rate
  analysis of cell-free massive {MIMO-NOMA} with three linear precoders,''
  \emph{IEEE Trans. Commun.}, vol.~68, no.~6, pp. 3480--3494, Jun. 2020.

\bibitem{nayebi2016performance}
E.~Nayebi, A.~Ashikhmin, T.~L. Marzetta, and B.~D. Rao, ``Performance of
  cell-free massive {MIMO} systems with {MMSE} and {LSFD} receivers,''
  \emph{Proc. IEEE ACSSC}, pp. 203--207, 2016.

\bibitem{nikbakht2019uplink}
R.~Nikbakht and A.~Lozano, ``Uplink fractional power control and downlink power
  allocation for cell-free networks,'' \emph{IEEE Wireless Commun. Lett.},
  vol.~9, no.~6, pp. 774--777, Jun. 2020.

\bibitem{bjornson2017massive}
E.~Bj{\"o}rnson, J.~Hoydis, and L.~Sanguinetti, ``Massive {MIMO} networks:
  Spectral, energy, and hardware efficiency,'' \emph{Foundations and
  Trends{\textregistered} in Signal Processing}, vol.~11, no. 3-4, pp.
  154--655, 2017.

\bibitem{Kay1996Fundamentals}
{S. M. Kay}, ``Fundamentals of statistical signal processing: Estimation
  theory,'' \emph{Prentice Hall}, 1993.

\bibitem{Marzetta2016Fundamentals}
T.~L. Marzetta, E.~G. Larsson, H.~Yang, and H.~Q. Ngo, \emph{Fundamentals of
  Massive {MIMO}}. Cambridge University
  Press, 2016.

\bibitem{Medard2000effect}
{M. Medard}, ``The effect upon channel capacity in wireless communications of
  perfect and imperfect knowledge of the channel,'' \emph{IEEE Trans. Inf.
  Theory}, vol.~46, no.~3, pp. 933--946, May 2000.

\bibitem{Zhang2020Cell-Free}
J.~{Zhang}, J.~{Zhang}, and B.~{Ai}, ``Cell-free massive {MIMO} with
  low-resolution {ADCs} over spatially correlated channels,'' in \emph{Proc.
  IEEE ICC}, Jun. 2020, pp. 1--7.

\bibitem{Jindal2011Multi}
{N. Jindal, J. G. Andrews, and S. Weber}, ``Multi-antenna communication in ad
  hoc networks: Achieving {MIMO} gains with {SIMO} transmission,'' \emph{IEEE
  Trans. Commun.}, vol.~59, no.~2, pp. 529--540, Feb. 2011.

\bibitem{Veetil2015Performance}
{S. T. Veetil, K. Kuchi, and R. K. Ganti}, ``Performance of {PZF} and {MMSE}
  receivers in cellular networks with multi-user spatial multiplexing,''
  \emph{IEEE Trans. Wireless Commun.}, vol.~14, no.~9, pp. 4867--4878, Sep.
  2015.

\bibitem{Fang2017Coverage}
{C. Fang, B. Makki, and T. Svensson}, ``Coverage analysis for millimeter wave
  uplink cellular networks with partial zero-forcing receivers,'' \emph{Proc.
  WiOpt}, pp. 1--6, May 2017.

\bibitem{Peel2005vector}
{C. Peel, B. Hochwald, and A. Swindlehurst}, ``A vector-perturbation technique
  for near-capacity multiantenna multiuser communication-Part {I}: Channel
  inversion and regularization,'' \emph{IEEE Trans. Commun.}, vol.~53, no.~1,
  pp. 195--202, Jan. 2005.

\bibitem{[R1]}
S.~{Wagner}, R.~{Couillet}, M.~{Debbah}, and D.~T.~M. {Slock}, ``Large system
  analysis of linear precoding in correlated {MISO} broadcast channels under
  limited feedback,'' \emph{IEEE Trans. Inf. Theory}, vol.~58, no.~7, pp.
  4509--4537, Jul. 2012.

\bibitem{[R2]}
H.~{Huh}, A.~M. {Tulino}, and G.~{Caire}, ``Network {MIMO} with linear
  zero-forcing beamforming: Large system analysis, impact of channel
  estimation, and reduced-complexity scheduling,'' \emph{IEEE Trans. Inf.
  Theory}, vol.~58, no.~5, pp. 2911--2934, May 2012.

\bibitem{Tulino2004random}
{A. M. Tulino} and {S. Verd{\'u}}, ``Random matrix theory and wireless
  communications,'' \emph{Foundations and Trends in Communications and
  Information Theory}, vol.~1, no.~1, pp. 1--182, 2004.

\bibitem{ozdogan2019massive}
{\"O}.~{\"O}zdogan, E.~Bj{\"o}rnson, and E.~G. Larsson, ``Massive {MIMO} with
  spatially correlated {Rician} fading channels,'' \emph{IEEE Trans. Commun.},
  vol.~67, no.~5, pp. 3234--3250, May 2019.

\end{thebibliography}
\end{document}